\documentclass[aps,twocolumn,showpacs,prl,amsmath,amssymb,floatfix,superscriptaddress]{revtex4-1}
\newcommand{\be}{\begin{equation}}
\newcommand{\ee}{\end{equation}}
\newcommand{\bea}{\begin{eqnarray}}
\newcommand{\eea}{\end{eqnarray}}
\usepackage{color}
\usepackage{graphicx}% Include figure files
\usepackage{dcolumn}% Align table columns on decimal point
\usepackage{bm}% bold math
\usepackage{array}
\usepackage{float}
\usepackage{supertabular}
\usepackage{longtable}
\usepackage{mathrsfs}
\usepackage{txfonts}
\usepackage{wasysym}

\begin{document}

\title{Emergent Supersymmetry from Strongly Interacting Majorana Zero Modes}

\author{Armin Rahmani}
\affiliation{Department of Physics and Astronomy and 
Quantum Matter Institute, University of British Columbia, Vancouver, British Columbia, Canada V6T 1Z4}

\author{Xiaoyu Zhu}
\affiliation{Department of Physics and Astronomy and 
Quantum Matter Institute, University of British Columbia, Vancouver, British Columbia, Canada V6T 1Z4}
\affiliation{National Laboratory of Solid State Microstructures and Department of Physics, Nanjing University - Nanjing 210093, China}

\author{Marcel Franz}
\affiliation{Department of Physics and Astronomy and 
Quantum Matter Institute, University of British Columbia, Vancouver, British Columbia, Canada V6T 1Z4}

\author{Ian Affleck}
\affiliation{Department of Physics and Astronomy and 
Quantum Matter Institute, University of British Columbia, Vancouver, British Columbia, Canada V6T 1Z4}

\date{\today}

\begin{abstract}

We show that a strongly interacting chain of Majorana zero modes exhibits a supersymmetric quantum critical point corresponding to the $c={7\over 10}$ tricritical Ising model, which separates a critical phase in the Ising universality class from a supersymmetric massive phase. We verify our predictions with numerical density-matrix-renormalization-group computations and determine the consequences for tunnelling experiments.

\end{abstract}
 
\maketitle

%
%The standard model of particle physics provides an accurate description of elementary particles and interactions up to a certain energy scale. It is, however, incomplete.

 Supersymmetry (SUSY) is a theoretical framework, which predicts a fermionic superpartner for every bosonic elementary particle, explaining long-standing puzzles (see, e.g., Ref.~\cite{Martin1997} and the references therein). The experimental verification of SUSY, however, has remained elusive. Another important milestone of physics is Majorana's prediction of fermionic particles that are their own antiparticle. Neutrinos were Majorana's original candidate for these so-called Majorana fermions, but their experimental status remains unclear~\cite{Wilczek2009, Elliott2015}.
%
% Interestingly, SUSY predicts the existence of other Majorana fermions such as the thus-far unobserverd photinos~\cite{Wilczek2009, Elliott2015} (spin-$1\over 2$ superpartners to photons). 

While both Majorana fermions and SUSY are yet to be observed in high energy physics, Majorana fermions are predicted to emerge as \textit{collective} excitations in many-body systems of electrons~\cite{Read2000,Kitaev2001, Stern2008, Nayak2008,Alicea2012,Fu2008,Lutchyn2010,Oreg2010}. Of particular interest are unpaired Majorana fermions, localized to topological defects such as vortices or domain walls, that occur at zero energy. These Majorana zero modes (MZMs) provide a promising candidate platform for topological quantum computing~\cite{Nayak2008}.
There has been significant recent experimental progress toward the detection of the condensed-matter incarnations of MZMs~\cite{Mourik2012,Das2012,Deng2012,Rokhinson2012,Finck2013,Hart2014,Nadj-Perge2014}. \textit{Strongly interacting Majoranas} may serve as building blocks for novel phases of matter, which remain relatively unexplored~\cite{Hassler2012,Terhal2012,Thomale2013,Kells2014,Chiu2015,Chiu2015b}.

{On the other hand, there have been few works on the realization of SUSY in condensed matter physics~\cite{Fendley2003,Feiguin2007,Huijse2008,Yu2010,Bauer2013,Huijse2015,Jian2015}. The canonical example of emergent SUSY in statistical physics is the tricritical Ising (TCI) model in $(1+1)$ dimensions~\cite{Friedan1984,Friedan1985,Qiu1986}. This model is is the second simplest unitary minimal conformal field theory (CFT) in $(1+1)$ dimensions. It has central charge $c={7\over 10}$ (Ising model with $c={1\over 2}$ being the simplest). It is also the only such CFT that exhibits SUSY.
% representing the universality class of the Landau-Ginzburg (LG) $\Phi^6$ theory at its tricritical point. (The simplest minimal model is the Ising model with $c={1\over 2}$, which represents the universality class of the $\Phi^4$ theory at its critical point.)

Similar to the Ising model, the TCI CFT has two realizations~\cite{Friedan1984}: (i) spin models such as the Blume-Capel model~\cite{Blume1966,Capel1966,Alcaraz1985}, in which, all local operators are bosonic, and (ii) fermionic models, in which both fermionic and bosonic local operators are present. Systems with Majorana fermions as local degrees of freedom provide promising candidates for realizing the fermionic models of TCI CFT~\cite{Zamolodchikov1991}. The local operator content of the TCI CFT, which determines the experimentally accessible correlation functions,  is directly related to the finite-size spectrum with periodic (antiperiodic) boundary conditions for the spin (fermionic) model.

%The Ising and TCI models have two 
%$Z_2$ symmetries - one which changes the sign of the magnetic (spin) operators and another one called Kramers-Wannier 
%duality which interchanges high temperature and low temperature phases of the classical model.

% When both these 
%symmetries are present, there is still one relevant operator allowed by symmetry which destablizies the TCI, and must 
%be fine-tuned to zero. 

In the TCI CFT, there is only one relevant operator allowed by symmetry (independent of the realization), which must be fine-tuned to zero. This relevant operator can destabilize the TCI CFT, causing a phase transition either to a doubly degenerate gapped phase or the Ising CFT phase (depending on the sign of the corresponding coupling constant).
%In the classical Ising model this corresponds to site dilution. 
This operator preserves the supersymmetry of the Hamiltonian~\cite{Friedan1984}. 
%One sign of the corresponding coupling constant drives the system 
%into the Ising CFT phase, whereas the other sign drives the system into a doubly degenerate gapped phase. 
SUSY is thus preserved in the 
gapped phase but spontaneously broken in the gapless Ising phase~\footnote{Right at the TCI point, SUSY is also spontaneously broken for a finite system with, say, periodic boundary conditions~\cite{Friedan1985}. The breaking of SUSY appears in $1/L$ corrections to the spectrum, while the operator content still reflects the superconformal structure of the CFT.}. While  nonsupersymmetric irrelevant operators could perturb the supersymmetric spectrum deep inside the gapped phase, in the vicinity of the critical point, the gapped phase is supersymmetric, with possible experimental signatures in tunneling experiments.

\begin{figure}[center]
	\includegraphics[width=7.5cm]{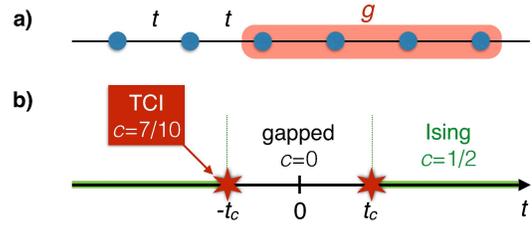}
	\caption{(a) The model with nearest-neighbor hopping and interactions between four nearest Majoranas. (b) The phase diagram of Hamiltonian~\eqref{eq:first} for $g=1$ as a function of $t$. }
	\label{fig:1}
\end{figure}

%
%Few microscopic models are known to realize the TCI CFT.
%The Blume-Capel model~\cite{Blume1966,Capel1966,Alcaraz1985} is a spin-$1$ chain (with fully bosonic local degrees of freedom), which realizes the Ramond sector of the theory upon tuning \textit{two} coupling constants in the Hamiltonian (this fine tuning makes it difficult to access the critical point in experiments).

Finding experimental realizations of the TCI CFT is of great interest. Important progress was made recently by constructing models of coupled bosonic degrees of freedom and fermionic Majorana modes, with the intuition that these local bosonic degrees of freedom can serve as superpartners to the Majoranas~\cite{Grover2014}. Purely fermionic field theories can also give rise to the TCI CFT~\cite{Kastor1989,Zamolodchikov1991}. Here we show that indeed the \textit{simplest lattice model} of interacting Majorana fermions, which may be realized in a superconducting vortex lattice and has clear experimental signatures in scanning tunneling microscopy (STM) experiments, gives rise to the TCI CFT and emergent SUSY. Similar to Ref.~\cite{Grover2014}, our model has the advantage that the TCI critical point can be reached by tuning only \textit{one} parameter (thanks to translation invariance of the model). The two models, however, differ in an important aspect. The model of Ref.~\cite{Grover2014} can be thought of as a model of interacting Majoranas upon integrating out the bosons. However, at the critical point, these boson-mediated interactions have a long-range character, whereas our interactions are strictly local.

}

We begin by writing the Hamiltonian
\begin{equation}\label{eq:first}
H=it\sum_{j}\gamma_{j}\gamma_{j+1}+g\sum_j\gamma_{j}\gamma_{j+1}\gamma_{j+2}\gamma_{j+3},
\end{equation}
where $\gamma_j=\gamma^\dagger_j$ (with $\{\gamma_i,\gamma_j\}=2\delta_{ij}$) is the annihilation (and creation) operator for a MZM at position $j$ in a one-dimensional lattice. This model can describe a vortex lattice in a narrow strip of a two dimensional topological superconductor~\cite{Chiu2015}. Throughout this paper, we focus on $g>0$, for which supersymmetric phases emerge (setting $g=1$ without loss of generality) \footnote{The sign of the hopping amplitude is unimportant as it can be changed via the transformation $\gamma_{j}\to (-1)^j\gamma_{j}$.}. Positive $g$ corresponds to attractive interactions between the underlying Dirac fermions, which may indeed appear in the presence of superconductivity.
The model exhibits a rich and complex phase diagram for $g<0$, which is discussed elsewhere~\cite{Milsted2015,Rahmani2015}.

Our main result is the phase diagram shown in Fig.~\ref{fig:1}. For $|t|>t_c$, the system is described by the Ising CFT with central charge $c={1\over 2}$, while for $0<|t|<t_c$, we have a gapped phase with broken symmetry. At $|t|=t_c$ the system realizes the $c={7\over 10}$ TCI model. As we will see, $t_c$ is extremely small (relative to the interaction strength $g$). However, the regime of strong interactions is accessible in experiments due to a \textit{chiral symmetry} (at chemical potential $\mu=0$ in vortex realization of Majoranas), which forbids hopping processes $it\gamma_j\gamma_{j'}$~\cite{Teo2010,Chiu2015}. By tuning $\mu$, we can then make $t$ arbitrarily small without changing the interactions.

% Two gapped phases with a dominant dimerization pattern are separated by a quantum phase transition at the translationally invariant line $g_1=g_2$. The quantum phase transition is first order (belongs to the Ising universality class) for $|t|<t_c$ ($|t|>t_c$). Precisely at $t_c$, the system realizes a tricritical Ising quantum critical point.

It is convenient \footnote{This is convenient because translation symmetry is spontaneously broken at strong coupling. } for the analysis of the problem to break the translation invariance of the system and write a more general Hamiltonian
\begin{equation}\label{eq:stag}
\begin{split}
H=&it_1\sum_{j}\alpha_{j}\beta_{j}
+it_2\sum_{j}\beta_{j}\alpha_{j+1}+
g_1\sum_j\alpha_{j}\beta_{j}\alpha_{j+1}\beta_{j+1}\\
&+g_2\sum_j\beta_{j}\alpha_{j+1}\beta_{j+1}\alpha_{j+2},
\end{split}
\end{equation}
where $\alpha_j\equiv \gamma_{2j}$ and $\beta_j\equiv \gamma_{2j+1}$. 
Each pair of Majoranas can be written in terms of one Dirac fermion $c_j=(\alpha_j+i\beta_j)/2$.
The product of two Majoranas is then related to the occupation number of a Dirac fermion through
$
i\alpha_{j}\beta_j=2n_j-1$,
where $n_j=c^\dagger_jc_j$. Note that the pairing of MZMs into Dirac fermions is rather arbitrary and we could have introduced another set of Dirac fermions $d_j=(\beta_j+i\alpha_{j+1})/2$.

In the limit of of $t\rightarrow\infty$ ($g/t \rightarrow 0 $), we see from Eq.~\eqref{eq:first} that the system is described by a free massless Majorana theory corresponding to the critical phase of the transverse field Ising model:
% the mapping to spin chain \eqref{eq:spin_chain} (with $t_1=t_2$) immediately implies that system must be in the critical phase of the transverse field Ising model. The low-energy effective theory of the Ising critical point is described by free massless Majorana fermions:
\begin{equation}\label{eq:effective}
H_t\approx iv\int dx \left(\gamma_R\partial_x{\gamma_R }-\gamma_L\partial_x{\gamma_L}\right),
\end{equation}
where $\alpha=2(\gamma_R+\gamma_L)$, $\beta=2(\gamma_R-\gamma_L)$, and $v$ is a velocity related to the renormalized hopping $t$.
% Note that the hopping term gives a Hamiltonian density
%$it\left(\beta \partial_x{\alpha }+\alpha \partial_x\beta\right)$.
 No mass term $im\int dx \gamma_R\gamma_L$ is present if we have translation symmetry.
 To incorporate the interactions into the effective theory, we Taylor expand the Majorana fields and obtain
$H_g=-256g\int dx \gamma_L\left(\partial_x\gamma_L\right)\gamma_R\left(\partial_x\gamma_R\right)+\cdots$,
where the dots indicate terms of third and higher order in derivatives. From simple power counting, we find that the perturbation above is irrelevant in the renormalization-group sense
% (interactions for $g_1=g_2$ do not generate a mass term due to translation invariance). 
 This implies that the Ising critical phase should extend \textit{at least }to a finite $g/t$.~\cite{Kastor1989}

\begin{figure}[center]
	\includegraphics[width=8cm]{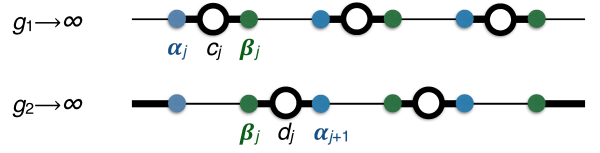}
	\caption{The symmetry broken states at strong coupling. The thick lines indicate pairs of
Majorana operators combined to form Dirac operators. The empty
circles symbolize that the resulting Dirac levels are empty. For both of the states shown, there is another degenerate state, where are all Dirac levels are filled.}
	\label{fig:2}
\end{figure}

 {Due to the large value of $g/t_c$, the strong coupling limit of the Hamiltonian provides a good qualitative understanding of the gapped phase.} It can be understood in terms of the occupation numbers of the $c$ and $d$ Dirac fermions as shown in Fig.~\ref{fig:2}. First, we consider the case of $t=0$. A dominant positive $g_1$ ($g_2$) gives ferromagnetic [all empty or all occupied] states for the occupation number of the $c$ ($d$) fermions. The phase transition at $t=0$ and $g_1=g_2$ (between phases with dominant $g_1$ and $g_2$) is expected to be \textit{first order} so (i) all of these ferromagnetic states are present in the ground state manifold of Hamiltonian~\eqref{eq:first} for $t=0$ and (ii) there is a gap to excitations. As discussed in the Supplemental Material (SM)~\footnote{See Supplemental Material [url], which includes Ref. \cite{Mahan}}, the presence of a first-order transition for $g_1=g_2$ and $t=0$ follows from the connection of our model to spin chains~\cite{Selke1988} and in particular a generalized Ising spin chains with multispin interactions, which bears a strong similarity to the 8-state Potts model~\cite{Turban1982,Penson1982,Alcaraz1986,Blote1986}. 
{Despite being a small perturbation, the hopping terms (depending on their sign) lift the degeneracy between the all-occupied and all-empty states leading to a doubly degenerate (instead of 4-fold) gapped phase for $0<|t|\ll g$.}

The analytical arguments for a gapped phase at strong coupling are not rigorous. However, if the strong coupling phase is a doubly degenerate gapped phase, and assuming there is only one phase transition between the Ising phase and this gapped phase, then the TCI CFT is the most natural theory of this phase transition~\cite{Note4}. The $c=1/2$ Ising ($c=0$ gapped) phase can be thought of as the line of a 2nd-order (1st-order) transition at $t_1=t_2$ ($g_1=g_2$) in a regime dominated by hopping (interaction). Similar to the Ising model with vacancies~\cite{Cardy}, the critical point, at which the 2nd-order transition changes to 1st-order transition, naturally corresponds (at the mean-field level) to the $\Phi^6$ Landau-Ginzburg theory at its tricritical point, whose universality class is given by the TCI CFT.

This scenario needs numerical verification. As discussed below, we found that the picture is indeed correct and the value of $t_c$ in the phase diagram of Fig.~\ref{fig:1} is $t_c/g=0.00405$. This small value in turn implies a gapped phase whose shortest correlation length is thousands of lattice sites. Establishing the nature of the phases and determining the value of $t_c$ is therefore exceedingly challenging with most numerical diagnostics such as extrapolation of gaps and order parameters as well as entanglement entropy. Despite this, we found that universal ratios in the finite-size spectrum provide a powerful numerical diagnostic for determining the phase diagram even though the system sizes we are able to reach are significantly smaller than the correlation length of the gapped phase in our model.

Our evidence for the fermionic TCI CFT is the excellent agreement between the theoretical predictions for several universal ratios  at the critical point shown in the table below, and the numerically computed values of these ratios. 
{\renewcommand{\arraystretch}{1.4}
\begin{center}
\begin{tabular}{c|c|c|c|c|c}
CFT& $c$ & ~${E^{\rm odd}_{{\rm A}, 0}-E^{\rm even}_{{\rm A}, 0}\over E^{\rm even}_{{\rm A}, 1}-E^{\rm even}_{{\rm A}, 0} }$~  & ~${E^{\rm even}_{{\rm P}, 0}-E^{\rm even}_{{\rm A}, 0}\over E^{\rm even}_{{\rm A}, 1}-E^{\rm even}_{{\rm A}, 0} }$~&~${E^{\rm even}_{{\rm P}, 1}-E^{\rm even}_{{\rm A}, 0}\over E^{\rm even}_{{\rm A}, 1}-E^{\rm even}_{{\rm A}, 0} }$~&~${E^{\rm even}_{{\rm A}, 0}-\epsilon_0 L\over E^{\rm even}_{{\rm A}, 1}-E^{\rm even}_{{\rm A}, 0} }$~ \\ 
\hline 
Ising & ${1\over 2}$ & ${1\over 2}$ &  ${1\over 8}$ &  ${1\over 4}$&  ${1\over 8}$\\ 
\hline\vspace{1mm} 
~TCI~ & ${7\over 10}$ & ${7\over 2}$& ${3\over 8}$  &${35\over 8}$ & ${7\over 24}$
\end{tabular} 
\end{center}
The subscripts A and P respectively indicate antiperiodic (APBC) and periodic (PBC) boundary conditions, $E_0$ and $E_1$ represent the energy of the ground state and the first excited state in a given fermion-parity sector (denoted by the superscripts even and odd), and $\epsilon_0$ is the thermodynamic-limit energy density in the ground state. The fourth universal ratio we use provides direct access to central charge $c$ through the general finite-size dependence of the ground-state energy (which is in the even parity sector for APBC): $E^{\rm even}_{{\rm A}, 0}=\epsilon_0 L-{2\pi v\over L}{c\over 12}$, where $L$ is the length of the system.
\footnote{We can accurately extract $E^{\rm even}_{{\rm A}, 0}-\epsilon_0 L$ from the numerics by a linear fit of the energy density $E^{\rm even}_{{\rm A}, 0}/L$ to $1/ L^2$.}.

We briefly outline the derivation of the above results based on the relationship between the operator content and the finite-size spectrum of the two CFTs. The results can be obtained from the formalism developed in Refs.~\cite{Cappelli1987,Lassig1991} as discussed in the SM~\cite{Note4}. We start with the Ising model, for which the predictions can be verified exactly in a free-fermion model. The Ising CFT has three primary fields $\mathbb{I}$ (identity), $\sigma$ (spin), and $\epsilon$ (energy) with conformal dimensions $h= \bar{h}=0, {1\over 16},{1\over 2}$, respectively [a field with conformal dimension $(h, \bar{h})$ has scaling correlators $\langle \phi_{h,\bar{h}}(x,t) \phi_{h,\bar{h}}(0,0)\rangle=(x-vt)^{-2h}(x+vt)^{-2\bar{h}}$]. An integer (half-odd-integer) conformal spin $h-\bar{h}$ corresponds to a bosonic (fermionic) excitation. Now in a fermionic theory, we have APBC in the imaginary time direction, which implies that modular invariance~\cite{Cardy1986} can be most easily satisfied if we also impose APBC in the spatial direction. The analog of the fermionic model in the Ising case is the free-Majorana model of Eq.~\eqref{eq:effective}, which has the conformal towers $(\mathbb{I},\mathbb{I})$, $(\mathbb{I},\epsilon)$, $(\epsilon,\mathbb{I})$, and $(\epsilon,\epsilon)$ with APBC (due to modular invariance~\cite{Cardy1986} as shown in the SM~\cite{Note4}), while the analog of the spin model has only diagonal conformal towers with bosonic excitations: $(\mathbb{I},\mathbb{I})$, $(\sigma,\sigma)$, and $(\epsilon,\epsilon)$.
\begin{figure}[center]
	\includegraphics[width=8cm]{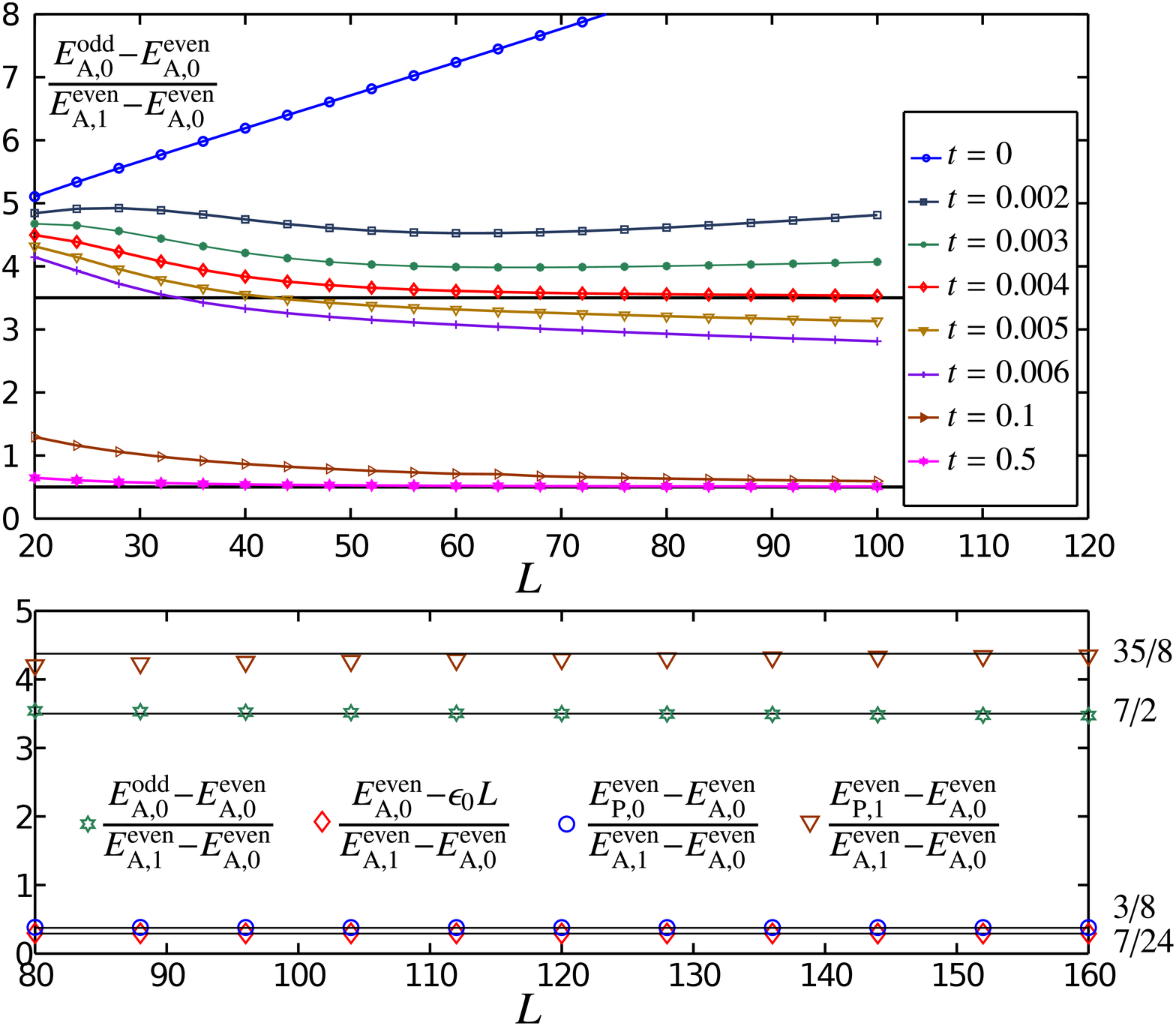}
	\caption{Top: The detection of the critical point through one of the universal ratios. Bottom: The values of four universal gap ratios at the tricritical point $t/g=0.00405$ as a function of $L$ (for systems with $2L$ Majoranas). The numerical results (data points) show excellent agreement with the CFT predictions (black lines at $7/2, 7/24, 3/8$, and $35/8$). The DMRG truncation errors are insignificant (up to 1000 states were kept in the computations).}
	\label{fig:3}
\end{figure}

In the fermionic model with APBC, the ground state and the first excited state of the even parity sector have operator content $(\mathbb{I},\mathbb{I})$ and $(\epsilon, \epsilon)$ respectively (note that the conformal spin vanishes implying even fermion parity). Similarly, the ground state in the odd parity sector is doubly degenerate with operator content $(\mathbb{I},\epsilon)$ and $(\epsilon, \mathbb{I})$. For the Ising CFT, we then obtain $E^{\rm even}_{{\rm A}, 1}-E^{\rm even}_{{\rm A}, 0}= {2\pi v\over L}({1\over 2}+{1\over 2})$ and $E^{\rm odd}_{{\rm A},0}-E^{\rm even}_{{\rm A}, 0}= {2\pi v\over L}(0+{1\over 2})$, which lead to the first universal ratio shown in the table above.

The TCI model has 6 primary fields $\mathbb{I}$, $\epsilon$, $\epsilon'$, $\epsilon''$, $\sigma$, and $\sigma'$, with scaling dimensions $h=\bar{h}=0, {1\over 10}, {3\over 5}, {3\over 2}, {3\over 80}, {7\over 16}$ respectively. Similar to the Ising case, we have $E^{\rm even}_{{\rm A}, 1}-E^{\rm even}_{{\rm A}, 0}= {2\pi v\over L}({1\over 10}+{1\over 10})$. However, in this case, $(\mathbb{I},\epsilon)$ does not appear in the modularly invariant conformal towers of the fermionic realization of TCI CFT (notice that a conformal spin of $1\over 10$ is neither an integer nor half integer). Here, the spin model also has 6 diagonal conformal towers, while, as shown in the SM the fermion model has 8 such towers with APBC, which include $(\epsilon,\epsilon')$, corresponding to the ground state of the odd sector with APBC~\cite{Note4}. This has a conformal spin ${3\over 5}-{1\over 10}={1\over 2}$. We then find $E^{\rm odd}_{{\rm A}, 0}-E^{\rm even}_{{\rm A}, 0}={2\pi v\over L}({3\over 5}+{1\over 10})$, leading to the first universal ratio in the table above. The spectrum with periodic boundary conditions is a bit more involved but can be similarly derived using CFT methods (see SM~\cite{Note4}).

 We numerically computed the four gap ratios above with the density-matrix-renormalization-group (DMRG) method. For the Ising and TCI CFT, the ratios above exhibit remarkable independence from the system size for large enough systems. On the other hand, in the gapped phase ($c=0$), at least one of the above ratios, namely ${E^{\rm odd}_{{\rm A}, 0}-E^{\rm even}_{{\rm A}, 0}\over E^{\rm even}_{{\rm A}, 1}-E^{\rm even}_{{\rm A}, 0} }$, grows with system size for large enough systems (it has a linear dependence on system size for $t=0$ as shown in Fig.~\ref{fig:3}). This gap ratio was used to detect the value of $t_c$ at the tricritical point. It plateaus at $7/2$ for $t_c$ and approaches the Ising value of $1/2$ for larger $t$. Having found the value of $t_c$, we then tested all four gap ratios for larger systems (see the bottom panel) and found excellent agreement with the theoretical predictions as seen in Fig.~\ref{fig:3}.

We finally discuss the experimental signatures of the TCI CFT. STM provides a powerful tool to probe local density of states. Tunneling into MZMs can effectively probe the critical exponent of the fermionic Green's function. We start with the Ising CFT~\eqref{eq:effective}. 
The nonvanishing fermionic correlators are $\langle \gamma_{L,R}(t,x),\gamma_{L,R}(0,0)\rangle={i\over 4\pi (vt\pm x+i\delta )}$, where $\delta$ is a positive infinitesimal number. The equal-time fermionic Green's function then decays as $1/x$. A closely related quantity is the tunneling current from an STM tip into a MZM, which goes as $I_{\rm I}\propto V$, where $V$ is the bias voltage.

In the TCI case, on the other hand, the leading fermionic operator $\chi$ corresponds to $(\epsilon,\epsilon')$ with $(h,\bar h)=(3/5,1/10)$, which gives $
\langle\chi(t,x)\chi(0,0)\rangle={i\over2\pi  (vt-x+i\delta )[(vt+i\delta )^2-x^2]^{1/5}}$,
leading to equal-time Green's functions, which decay as $|x|^{-7/5}$. The tunneling current then goes as~\cite{Note4}
\begin{equation}
I_{\rm TCI}\propto \hbox{sign}(V)|V|^{7/5}.
\end{equation}

In Fig.~\ref{fig:4}, we show the scaling behavior of the equal-time Green's function of our model computed for TCI ($t=t_c$) and Ising ($t=10^3\sim\infty$), where the predicted exponents are easily observed ($g=1$ in both cases). The only relevant operator (that induces a transition to the gapped phase from the TCI critical point) is $(\epsilon ',\epsilon ')$ with dimension $3/5+3/5=6/5$. We then expect a gap in the symmetry-broken phase that scales as $(t_c-t)^{5/4}$ near the critical point~\cite{Note4}.

As mentioned before, an important property of the relevant $(\epsilon ',\epsilon ')$ operator is that it is supersymmetric~\cite{Kastor1989,Lassig1991,Zamolodchikov1991}. Therefore the SUSY of the critical point should extend into the gapped phase at least in the vicinity of the critical point. In the gapped phase, the power-law dependence of the tunneling current on $V$ changes to  exponential dependence, from which the gap to the leading fermionic excitation can be extracted. SUSY implies that the leading bosonic excitation has the same gap as the leading fermionic one. It should be possible to experimentally determine this bosonic gap from Cooper-pair tunneling via a superconducting tip or other bosonic probes such as coupling to photons or phonons. 

Considering the effect of disorder on the rich physics of interacting Majoranas adds a new dimension to the problem~\cite{Cheng2009}: a recent manuscript, which appeared shortly after the present paper, examines the effects~\cite{Milsted2015}. {Our theory applies to a translationally invariant system. Experimentally, it is common to form Abrikisov vortex lattices with translation invariance due to energetic reasons. Spontaneous dimerization of the vortex lattice might occur which would indeed 
gap the system and destroy the TCI point.  However, even in that case, if the dimerization is weak, some signatures of the critical point survive in a crossover regime.}

\begin{figure}[center]
	\includegraphics[width=8cm]{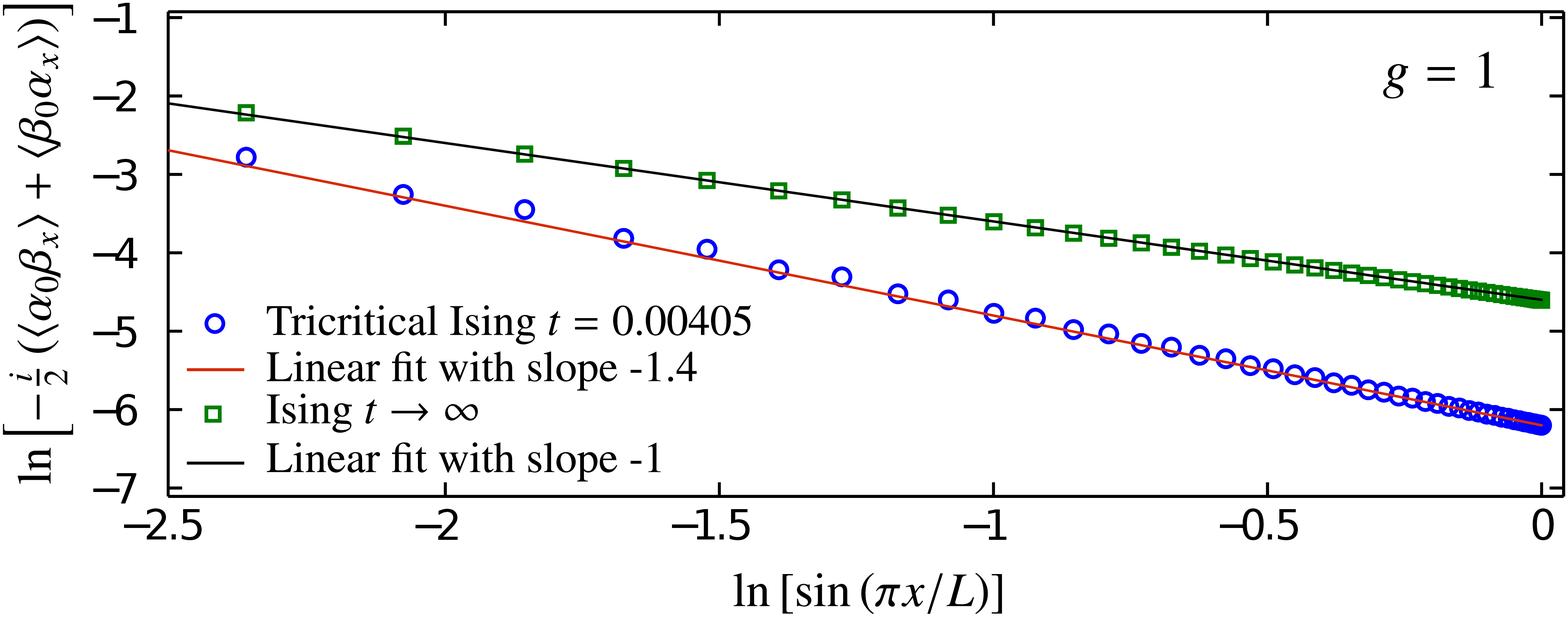}
	\caption{The scaling behavior of the fermionic Green's function (obtained with DMRG keeping 800 states) for the Ising and the TCI CFTs observed in our microscopic model.}
	\label{fig:4}
\end{figure}

%Finally, we comment on the effects of disorder on our system. The Ising phase at least in weak coupling regime is stable again disorder. states are expected to remain extended at zero energy~\cite{Eggarter1978}. This was also numerically verified in a recent manuscript~\cite{Milsted2015}. Also deep in the gapped phase we don not expect weak disorder to have any important effects. Whether the TCI critical point remains stable against disorder is an important question, which goes beyond the scope of the present paper.

In summary, we studied the phase diagram of the simplest model of strongly interacting Majorana zero modes in one dimension for attractive underlying interactions (which may be realized in the presence of superconductivity). Supported by extensive numerical calculations, we demonstrated that our model provides one of the few examples of emergent spacetime SUSY in condensed matter physics and the first lattice model realization of TCI SUSY in a purely fermionic system with local interactions (a different type of SUSY has been predicted in a lattice model with 6-fermion interactions~\cite{Fendley2003}). The vortex lattice experimental realization of our model fosters the thus far elusive observation of SUSY, with clear signatures in the behavior of the tunneling current into a Majorana mode.

\acknowledgements{We thank T. Grover for helpful discussions. This work was supported by NSERC (IA, MF, and AR), CIfAR (IA and MF), Max Planck-UBC Centre for Quantum Materials (IA, MF, and AR) and China Scholarship Council (XZ).}
\bibliography{majorana.bib}

%merlin.mbs apsrev4-1.bst 2010-07-25 4.21a (PWD, AO, DPC) hacked
%Control: key (0)
%Control: author (8) initials jnrlst
%Control: editor formatted (1) identically to author
%Control: production of article title (-1) disabled
%Control: page (0) single
%Control: year (1) truncated
%Control: production of eprint (0) enabled
\begin{thebibliography}{58}%
\makeatletter
\providecommand \@ifxundefined [1]{%
 \@ifx{#1\undefined}
}%
\providecommand \@ifnum [1]{%
 \ifnum #1\expandafter \@firstoftwo
 \else \expandafter \@secondoftwo
 \fi
}%
\providecommand \@ifx [1]{%
 \ifx #1\expandafter \@firstoftwo
 \else \expandafter \@secondoftwo
 \fi
}%
\providecommand \natexlab [1]{#1}%
\providecommand \enquote  [1]{``#1''}%
\providecommand \bibnamefont  [1]{#1}%
\providecommand \bibfnamefont [1]{#1}%
\providecommand \citenamefont [1]{#1}%
\providecommand \href@noop [0]{\@secondoftwo}%
\providecommand \href [0]{\begingroup \@sanitize@url \@href}%
\providecommand \@href[1]{\@@startlink{#1}\@@href}%
\providecommand \@@href[1]{\endgroup#1\@@endlink}%
\providecommand \@sanitize@url [0]{\catcode `\\12\catcode `\$12\catcode
  `\&12\catcode `\#12\catcode `\^12\catcode `\_12\catcode `\%12\relax}%
\providecommand \@@startlink[1]{}%
\providecommand \@@endlink[0]{}%
\providecommand \url  [0]{\begingroup\@sanitize@url \@url }%
\providecommand \@url [1]{\endgroup\@href {#1}{\urlprefix }}%
\providecommand \urlprefix  [0]{URL }%
\providecommand \Eprint [0]{\href }%
\providecommand \doibase [0]{http://dx.doi.org/}%
\providecommand \selectlanguage [0]{\@gobble}%
\providecommand \bibinfo  [0]{\@secondoftwo}%
\providecommand \bibfield  [0]{\@secondoftwo}%
\providecommand \translation [1]{[#1]}%
\providecommand \BibitemOpen [0]{}%
\providecommand \bibitemStop [0]{}%
\providecommand \bibitemNoStop [0]{.\EOS\space}%
\providecommand \EOS [0]{\spacefactor3000\relax}%
\providecommand \BibitemShut  [1]{\csname bibitem#1\endcsname}%
\let\auto@bib@innerbib\@empty
%</preamble>
\bibitem [{\citenamefont {Martin}()}]{Martin1997}%
  \BibitemOpen
  \bibfield  {author} {\bibinfo {author} {\bibfnamefont {S.~P.}\ \bibnamefont
  {Martin}},\ }\href@noop {} {}\Eprint {http://arxiv.org/abs/hep-ph/9709356}
  {arXiv:hep-ph/9709356} \BibitemShut {NoStop}%
\bibitem [{\citenamefont {Wilczek}(2009)}]{Wilczek2009}%
  \BibitemOpen
  \bibfield  {author} {\bibinfo {author} {\bibfnamefont {F.}~\bibnamefont
  {Wilczek}},\ }\href@noop {} {\bibfield  {journal} {\bibinfo  {journal} {Nat.
  Phys.}\ }\textbf {\bibinfo {volume} {5}},\ \bibinfo {pages} {614} (\bibinfo
  {year} {2009})}\BibitemShut {NoStop}%
\bibitem [{\citenamefont {Elliott}\ and\ \citenamefont
  {Franz}(2015)}]{Elliott2015}%
  \BibitemOpen
  \bibfield  {author} {\bibinfo {author} {\bibfnamefont {S.~R.}\ \bibnamefont
  {Elliott}}\ and\ \bibinfo {author} {\bibfnamefont {M.}~\bibnamefont
  {Franz}},\ }\href@noop {} {\bibfield  {journal} {\bibinfo  {journal} {Rev.
  Mod. Phys.}\ }\textbf {\bibinfo {volume} {87}},\ \bibinfo {pages} {137}
  (\bibinfo {year} {2015})}\BibitemShut {NoStop}%
\bibitem [{\citenamefont {Read}\ and\ \citenamefont {Green}(2000)}]{Read2000}%
  \BibitemOpen
  \bibfield  {author} {\bibinfo {author} {\bibfnamefont {N.}~\bibnamefont
  {Read}}\ and\ \bibinfo {author} {\bibfnamefont {D.}~\bibnamefont {Green}},\
  }\href@noop {} {\bibfield  {journal} {\bibinfo  {journal} {Phys. Rev. B}\
  }\textbf {\bibinfo {volume} {61}},\ \bibinfo {pages} {10267} (\bibinfo {year}
  {2000})}\BibitemShut {NoStop}%
\bibitem [{\citenamefont {Kitaev}(2001)}]{Kitaev2001}%
  \BibitemOpen
  \bibfield  {author} {\bibinfo {author} {\bibfnamefont {A.~Y.}\ \bibnamefont
  {Kitaev}},\ }\href@noop {} {\bibfield  {journal} {\bibinfo  {journal}
  {Physics-Uspekhi}\ }\textbf {\bibinfo {volume} {44}},\ \bibinfo {pages} {131}
  (\bibinfo {year} {2001})}\BibitemShut {NoStop}%
\bibitem [{\citenamefont {Stern}(2008)}]{Stern2008}%
  \BibitemOpen
  \bibfield  {author} {\bibinfo {author} {\bibfnamefont {A.}~\bibnamefont
  {Stern}},\ }\href@noop {} {\bibfield  {journal} {\bibinfo  {journal} {Ann.
  Phys.}\ }\textbf {\bibinfo {volume} {323}},\ \bibinfo {pages} {204} (\bibinfo
  {year} {2008})}\BibitemShut {NoStop}%
\bibitem [{\citenamefont {Nayak}\ \emph {et~al.}(2008)\citenamefont {Nayak},
  \citenamefont {Simon}, \citenamefont {Stern}, \citenamefont {Freedman},\ and\
  \citenamefont {Das~Sarma}}]{Nayak2008}%
  \BibitemOpen
  \bibfield  {author} {\bibinfo {author} {\bibfnamefont {C.}~\bibnamefont
  {Nayak}}, \bibinfo {author} {\bibfnamefont {S.~H.}\ \bibnamefont {Simon}},
  \bibinfo {author} {\bibfnamefont {A.}~\bibnamefont {Stern}}, \bibinfo
  {author} {\bibfnamefont {M.}~\bibnamefont {Freedman}}, \ and\ \bibinfo
  {author} {\bibfnamefont {S.}~\bibnamefont {Das~Sarma}},\ }\href@noop {}
  {\bibfield  {journal} {\bibinfo  {journal} {Rev. Mod. Phys.}\ }\textbf
  {\bibinfo {volume} {80}},\ \bibinfo {pages} {1083} (\bibinfo {year}
  {2008})}\BibitemShut {NoStop}%
\bibitem [{\citenamefont {Alicea}(2012)}]{Alicea2012}%
  \BibitemOpen
  \bibfield  {author} {\bibinfo {author} {\bibfnamefont {J.}~\bibnamefont
  {Alicea}},\ }\href@noop {} {\bibfield  {journal} {\bibinfo  {journal} {Rep.
  Prog. Phys.}\ }\textbf {\bibinfo {volume} {75}},\ \bibinfo {pages} {076501}
  (\bibinfo {year} {2012})}\BibitemShut {NoStop}%
\bibitem [{\citenamefont {Fu}\ and\ \citenamefont {Kane}(2008)}]{Fu2008}%
  \BibitemOpen
  \bibfield  {author} {\bibinfo {author} {\bibfnamefont {L.}~\bibnamefont
  {Fu}}\ and\ \bibinfo {author} {\bibfnamefont {C.~L.}\ \bibnamefont {Kane}},\
  }\href@noop {} {\bibfield  {journal} {\bibinfo  {journal} {Phys. Rev. Lett.}\
  }\textbf {\bibinfo {volume} {100}},\ \bibinfo {pages} {096407} (\bibinfo
  {year} {2008})}\BibitemShut {NoStop}%
\bibitem [{\citenamefont {Lutchyn}\ \emph {et~al.}(2010)\citenamefont
  {Lutchyn}, \citenamefont {Sau},\ and\ \citenamefont
  {Das~Sarma}}]{Lutchyn2010}%
  \BibitemOpen
  \bibfield  {author} {\bibinfo {author} {\bibfnamefont {R.~M.}\ \bibnamefont
  {Lutchyn}}, \bibinfo {author} {\bibfnamefont {J.~D.}\ \bibnamefont {Sau}}, \
  and\ \bibinfo {author} {\bibfnamefont {S.}~\bibnamefont {Das~Sarma}},\
  }\href@noop {} {\bibfield  {journal} {\bibinfo  {journal} {Phys. Rev. Lett.}\
  }\textbf {\bibinfo {volume} {105}},\ \bibinfo {pages} {077001} (\bibinfo
  {year} {2010})}\BibitemShut {NoStop}%
\bibitem [{\citenamefont {Oreg}\ \emph {et~al.}(2010)\citenamefont {Oreg},
  \citenamefont {Refael},\ and\ \citenamefont {von Oppen}}]{Oreg2010}%
  \BibitemOpen
  \bibfield  {author} {\bibinfo {author} {\bibfnamefont {Y.}~\bibnamefont
  {Oreg}}, \bibinfo {author} {\bibfnamefont {G.}~\bibnamefont {Refael}}, \ and\
  \bibinfo {author} {\bibfnamefont {F.}~\bibnamefont {von Oppen}},\ }\href@noop
  {} {\bibfield  {journal} {\bibinfo  {journal} {Phys. Rev. Lett.}\ }\textbf
  {\bibinfo {volume} {105}},\ \bibinfo {pages} {177002} (\bibinfo {year}
  {2010})}\BibitemShut {NoStop}%
\bibitem [{\citenamefont {Mourik}\ \emph {et~al.}(2012)\citenamefont {Mourik},
  \citenamefont {Zuo}, \citenamefont {Frolov}, \citenamefont {Plissard},
  \citenamefont {Bakkers},\ and\ \citenamefont {Kouwenhoven}}]{Mourik2012}%
  \BibitemOpen
  \bibfield  {author} {\bibinfo {author} {\bibfnamefont {V.}~\bibnamefont
  {Mourik}}, \bibinfo {author} {\bibfnamefont {K.}~\bibnamefont {Zuo}},
  \bibinfo {author} {\bibfnamefont {S.~M.}\ \bibnamefont {Frolov}}, \bibinfo
  {author} {\bibfnamefont {S.~R.}\ \bibnamefont {Plissard}}, \bibinfo {author}
  {\bibfnamefont {E.~P. a.~M.}\ \bibnamefont {Bakkers}}, \ and\ \bibinfo
  {author} {\bibfnamefont {L.~P.}\ \bibnamefont {Kouwenhoven}},\ }\href@noop {}
  {\bibfield  {journal} {\bibinfo  {journal} {Science}\ }\textbf {\bibinfo
  {volume} {336}},\ \bibinfo {pages} {1003} (\bibinfo {year}
  {2012})}\BibitemShut {NoStop}%
\bibitem [{\citenamefont {Das}\ \emph {et~al.}(2012)\citenamefont {Das},
  \citenamefont {Ronen}, \citenamefont {Most}, \citenamefont {Oreg},
  \citenamefont {Heiblum},\ and\ \citenamefont {Shtrikman}}]{Das2012}%
  \BibitemOpen
  \bibfield  {author} {\bibinfo {author} {\bibfnamefont {A.}~\bibnamefont
  {Das}}, \bibinfo {author} {\bibfnamefont {Y.}~\bibnamefont {Ronen}}, \bibinfo
  {author} {\bibfnamefont {Y.}~\bibnamefont {Most}}, \bibinfo {author}
  {\bibfnamefont {Y.}~\bibnamefont {Oreg}}, \bibinfo {author} {\bibfnamefont
  {M.}~\bibnamefont {Heiblum}}, \ and\ \bibinfo {author} {\bibfnamefont
  {H.}~\bibnamefont {Shtrikman}},\ }\href@noop {} {\bibfield  {journal}
  {\bibinfo  {journal} {Nat. Phys.}\ }\textbf {\bibinfo {volume} {8}},\
  \bibinfo {pages} {887} (\bibinfo {year} {2012})}\BibitemShut {NoStop}%
\bibitem [{\citenamefont {Deng}\ \emph {et~al.}(2012)\citenamefont {Deng},
  \citenamefont {Yu}, \citenamefont {Huang}, \citenamefont {Larsson},
  \citenamefont {Caroff},\ and\ \citenamefont {Xu}}]{Deng2012}%
  \BibitemOpen
  \bibfield  {author} {\bibinfo {author} {\bibfnamefont {M.~T.}\ \bibnamefont
  {Deng}}, \bibinfo {author} {\bibfnamefont {C.~L.}\ \bibnamefont {Yu}},
  \bibinfo {author} {\bibfnamefont {G.~Y.}\ \bibnamefont {Huang}}, \bibinfo
  {author} {\bibfnamefont {M.}~\bibnamefont {Larsson}}, \bibinfo {author}
  {\bibfnamefont {P.}~\bibnamefont {Caroff}}, \ and\ \bibinfo {author}
  {\bibfnamefont {H.~Q.}\ \bibnamefont {Xu}},\ }\href@noop {} {\bibfield
  {journal} {\bibinfo  {journal} {Nano Lett.}\ }\textbf {\bibinfo {volume}
  {12}},\ \bibinfo {pages} {6414} (\bibinfo {year} {2012})}\BibitemShut
  {NoStop}%
\bibitem [{\citenamefont {Rokhinson}\ \emph {et~al.}(2012)\citenamefont
  {Rokhinson}, \citenamefont {Liu},\ and\ \citenamefont
  {Furdyna}}]{Rokhinson2012}%
  \BibitemOpen
  \bibfield  {author} {\bibinfo {author} {\bibfnamefont {L.~P.}\ \bibnamefont
  {Rokhinson}}, \bibinfo {author} {\bibfnamefont {X.}~\bibnamefont {Liu}}, \
  and\ \bibinfo {author} {\bibfnamefont {J.~K.}\ \bibnamefont {Furdyna}},\
  }\href@noop {} {\bibfield  {journal} {\bibinfo  {journal} {Nat. Phys.}\
  }\textbf {\bibinfo {volume} {8}},\ \bibinfo {pages} {795} (\bibinfo {year}
  {2012})}\BibitemShut {NoStop}%
\bibitem [{\citenamefont {Finck}\ \emph {et~al.}(2013)\citenamefont {Finck},
  \citenamefont {Van~Harlingen}, \citenamefont {Mohseni}, \citenamefont
  {Jung},\ and\ \citenamefont {Li}}]{Finck2013}%
  \BibitemOpen
  \bibfield  {author} {\bibinfo {author} {\bibfnamefont {A.~D.~K.}\
  \bibnamefont {Finck}}, \bibinfo {author} {\bibfnamefont {D.~J.}\ \bibnamefont
  {Van~Harlingen}}, \bibinfo {author} {\bibfnamefont {P.~K.}\ \bibnamefont
  {Mohseni}}, \bibinfo {author} {\bibfnamefont {K.}~\bibnamefont {Jung}}, \
  and\ \bibinfo {author} {\bibfnamefont {X.}~\bibnamefont {Li}},\ }\href@noop
  {} {\bibfield  {journal} {\bibinfo  {journal} {Phys. Rev. Lett.}\ }\textbf
  {\bibinfo {volume} {110}},\ \bibinfo {pages} {126406} (\bibinfo {year}
  {2013})}\BibitemShut {NoStop}%
\bibitem [{\citenamefont {Hart}\ \emph {et~al.}(2014)\citenamefont {Hart},
  \citenamefont {Ren}, \citenamefont {Wagner}, \citenamefont {Leubner},
  \citenamefont {M{\"u}hlbauer}, \citenamefont {Br{\"u}ne}, \citenamefont
  {Buhmann}, \citenamefont {Molenkamp},\ and\ \citenamefont
  {Yacoby}}]{Hart2014}%
  \BibitemOpen
  \bibfield  {author} {\bibinfo {author} {\bibfnamefont {S.}~\bibnamefont
  {Hart}}, \bibinfo {author} {\bibfnamefont {H.}~\bibnamefont {Ren}}, \bibinfo
  {author} {\bibfnamefont {T.}~\bibnamefont {Wagner}}, \bibinfo {author}
  {\bibfnamefont {P.}~\bibnamefont {Leubner}}, \bibinfo {author} {\bibfnamefont
  {M.}~\bibnamefont {M{\"u}hlbauer}}, \bibinfo {author} {\bibfnamefont
  {C.}~\bibnamefont {Br{\"u}ne}}, \bibinfo {author} {\bibfnamefont
  {H.}~\bibnamefont {Buhmann}}, \bibinfo {author} {\bibfnamefont {L.~W.}\
  \bibnamefont {Molenkamp}}, \ and\ \bibinfo {author} {\bibfnamefont
  {A.}~\bibnamefont {Yacoby}},\ }\href@noop {} {\bibfield  {journal} {\bibinfo
  {journal} {Nat. Phys.}\ }\textbf {\bibinfo {volume} {10}},\ \bibinfo {pages}
  {638} (\bibinfo {year} {2014})}\BibitemShut {NoStop}%
\bibitem [{\citenamefont {Nadj-Perge}\ \emph {et~al.}(2014)\citenamefont
  {Nadj-Perge}, \citenamefont {Drozdov}, \citenamefont {Li}, \citenamefont
  {Chen}, \citenamefont {Jeon}, \citenamefont {Seo}, \citenamefont {MacDonald},
  \citenamefont {Bernevig},\ and\ \citenamefont {Yazdani}}]{Nadj-Perge2014}%
  \BibitemOpen
  \bibfield  {author} {\bibinfo {author} {\bibfnamefont {S.}~\bibnamefont
  {Nadj-Perge}}, \bibinfo {author} {\bibfnamefont {I.~K.}\ \bibnamefont
  {Drozdov}}, \bibinfo {author} {\bibfnamefont {J.}~\bibnamefont {Li}},
  \bibinfo {author} {\bibfnamefont {H.}~\bibnamefont {Chen}}, \bibinfo {author}
  {\bibfnamefont {S.}~\bibnamefont {Jeon}}, \bibinfo {author} {\bibfnamefont
  {J.}~\bibnamefont {Seo}}, \bibinfo {author} {\bibfnamefont {A.~H.}\
  \bibnamefont {MacDonald}}, \bibinfo {author} {\bibfnamefont {B.~A.}\
  \bibnamefont {Bernevig}}, \ and\ \bibinfo {author} {\bibfnamefont
  {A.}~\bibnamefont {Yazdani}},\ }\href@noop {} {\bibfield  {journal} {\bibinfo
   {journal} {Science}\ }\textbf {\bibinfo {volume} {346}},\ \bibinfo {pages}
  {602} (\bibinfo {year} {2014})}\BibitemShut {NoStop}%
\bibitem [{\citenamefont {Hassler}\ and\ \citenamefont
  {Schuricht}(2014)}]{Hassler2012}%
  \BibitemOpen
  \bibfield  {author} {\bibinfo {author} {\bibfnamefont {F.}~\bibnamefont
  {Hassler}}\ and\ \bibinfo {author} {\bibfnamefont {D.}~\bibnamefont
  {Schuricht}},\ }\href@noop {} {\bibfield  {journal} {\bibinfo  {journal} {New
  J. Phys.}\ }\textbf {\bibinfo {volume} {14}},\ \bibinfo {pages} {125018}
  (\bibinfo {year} {2014})}\BibitemShut {NoStop}%
\bibitem [{\citenamefont {Terhal}\ \emph {et~al.}(2012)\citenamefont {Terhal},
  \citenamefont {Hassler},\ and\ \citenamefont {DiVincenzo}}]{Terhal2012}%
  \BibitemOpen
  \bibfield  {author} {\bibinfo {author} {\bibfnamefont {B.~M.}\ \bibnamefont
  {Terhal}}, \bibinfo {author} {\bibfnamefont {F.}~\bibnamefont {Hassler}}, \
  and\ \bibinfo {author} {\bibfnamefont {D.~P.}\ \bibnamefont {DiVincenzo}},\
  }\href@noop {} {\bibfield  {journal} {\bibinfo  {journal} {Phys. Rev. Lett.}\
  }\textbf {\bibinfo {volume} {108}},\ \bibinfo {pages} {260504} (\bibinfo
  {year} {2012})}\BibitemShut {NoStop}%
\bibitem [{\citenamefont {Thomale}\ \emph {et~al.}(2013)\citenamefont
  {Thomale}, \citenamefont {Rachel},\ and\ \citenamefont
  {Schmitteckert}}]{Thomale2013}%
  \BibitemOpen
  \bibfield  {author} {\bibinfo {author} {\bibfnamefont {R.}~\bibnamefont
  {Thomale}}, \bibinfo {author} {\bibfnamefont {S.}~\bibnamefont {Rachel}}, \
  and\ \bibinfo {author} {\bibfnamefont {P.}~\bibnamefont {Schmitteckert}},\
  }\href@noop {} {\bibfield  {journal} {\bibinfo  {journal} {Phys. Rev. B}\
  }\textbf {\bibinfo {volume} {88}},\ \bibinfo {pages} {161103} (\bibinfo
  {year} {2013})}\BibitemShut {NoStop}%
\bibitem [{\citenamefont {Kells}\ \emph {et~al.}(2014)\citenamefont {Kells},
  \citenamefont {Lahtinen},\ and\ \citenamefont {Vala}}]{Kells2014}%
  \BibitemOpen
  \bibfield  {author} {\bibinfo {author} {\bibfnamefont {G.}~\bibnamefont
  {Kells}}, \bibinfo {author} {\bibfnamefont {V.}~\bibnamefont {Lahtinen}}, \
  and\ \bibinfo {author} {\bibfnamefont {J.}~\bibnamefont {Vala}},\ }\href@noop
  {} {\bibfield  {journal} {\bibinfo  {journal} {Phys. Rev. B}\ }\textbf
  {\bibinfo {volume} {89}},\ \bibinfo {pages} {075122} (\bibinfo {year}
  {2014})}\BibitemShut {NoStop}%
\bibitem [{\citenamefont {Chiu}\ \emph {et~al.}(2015)\citenamefont {Chiu},
  \citenamefont {Pikulin},\ and\ \citenamefont {Franz}}]{Chiu2015}%
  \BibitemOpen
  \bibfield  {author} {\bibinfo {author} {\bibfnamefont {C.-K.}\ \bibnamefont
  {Chiu}}, \bibinfo {author} {\bibfnamefont {D.~I.}\ \bibnamefont {Pikulin}}, \
  and\ \bibinfo {author} {\bibfnamefont {M.}~\bibnamefont {Franz}},\
  }\href@noop {} {\bibfield  {journal} {\bibinfo  {journal} {Phys. Rev. B}\
  }\textbf {\bibinfo {volume} {91}},\ \bibinfo {pages} {165402} (\bibinfo
  {year} {2015})}\BibitemShut {NoStop}%
\bibitem [{\citenamefont {Chiu}\ \emph {et~al.}()\citenamefont {Chiu},
  \citenamefont {Pikulin},\ and\ \citenamefont {Franz}}]{Chiu2015b}%
  \BibitemOpen
  \bibfield  {author} {\bibinfo {author} {\bibfnamefont {C.-K.}\ \bibnamefont
  {Chiu}}, \bibinfo {author} {\bibfnamefont {D.~I.}\ \bibnamefont {Pikulin}}, \
  and\ \bibinfo {author} {\bibfnamefont {M.}~\bibnamefont {Franz}},\
  }\href@noop {} {}\Eprint {http://arxiv.org/abs/1502.03432} {arXiv:1502.03432}
  \BibitemShut {NoStop}%
\bibitem [{\citenamefont {Fendley}\ \emph {et~al.}(2003)\citenamefont
  {Fendley}, \citenamefont {Schoutens},\ and\ \citenamefont
  {de~Boer}}]{Fendley2003}%
  \BibitemOpen
  \bibfield  {author} {\bibinfo {author} {\bibfnamefont {P.}~\bibnamefont
  {Fendley}}, \bibinfo {author} {\bibfnamefont {K.}~\bibnamefont {Schoutens}},
  \ and\ \bibinfo {author} {\bibfnamefont {J.}~\bibnamefont {de~Boer}},\
  }\href@noop {} {\bibfield  {journal} {\bibinfo  {journal} {Phys. Rev. Lett.}\
  }\textbf {\bibinfo {volume} {90}},\ \bibinfo {pages} {120402} (\bibinfo
  {year} {2003})}\BibitemShut {NoStop}%
\bibitem [{\citenamefont {Feiguin}\ \emph {et~al.}(2007)\citenamefont
  {Feiguin}, \citenamefont {Trebst}, \citenamefont {Ludwig}, \citenamefont
  {Troyer}, \citenamefont {Kitaev}, \citenamefont {Wang},\ and\ \citenamefont
  {Freedman}}]{Feiguin2007}%
  \BibitemOpen
  \bibfield  {author} {\bibinfo {author} {\bibfnamefont {A.}~\bibnamefont
  {Feiguin}}, \bibinfo {author} {\bibfnamefont {S.}~\bibnamefont {Trebst}},
  \bibinfo {author} {\bibfnamefont {A.~W.~W.}\ \bibnamefont {Ludwig}}, \bibinfo
  {author} {\bibfnamefont {M.}~\bibnamefont {Troyer}}, \bibinfo {author}
  {\bibfnamefont {A.}~\bibnamefont {Kitaev}}, \bibinfo {author} {\bibfnamefont
  {Z.}~\bibnamefont {Wang}}, \ and\ \bibinfo {author} {\bibfnamefont {M.~H.}\
  \bibnamefont {Freedman}},\ }\href@noop {} {\bibfield  {journal} {\bibinfo
  {journal} {Phys. Rev. Lett.}\ }\textbf {\bibinfo {volume} {98}},\ \bibinfo
  {pages} {160409} (\bibinfo {year} {2007})}\BibitemShut {NoStop}%
\bibitem [{\citenamefont {Huijse}\ \emph {et~al.}(2008)\citenamefont {Huijse},
  \citenamefont {Halverson}, \citenamefont {Fendley},\ and\ \citenamefont
  {Schoutens}}]{Huijse2008}%
  \BibitemOpen
  \bibfield  {author} {\bibinfo {author} {\bibfnamefont {L.}~\bibnamefont
  {Huijse}}, \bibinfo {author} {\bibfnamefont {J.}~\bibnamefont {Halverson}},
  \bibinfo {author} {\bibfnamefont {P.}~\bibnamefont {Fendley}}, \ and\
  \bibinfo {author} {\bibfnamefont {K.}~\bibnamefont {Schoutens}},\ }\href@noop
  {} {\bibfield  {journal} {\bibinfo  {journal} {Phys. Rev. Lett.}\ }\textbf
  {\bibinfo {volume} {101}},\ \bibinfo {pages} {146406} (\bibinfo {year}
  {2008})}\BibitemShut {NoStop}%
\bibitem [{\citenamefont {Yu}\ and\ \citenamefont {Yang}(2010)}]{Yu2010}%
  \BibitemOpen
  \bibfield  {author} {\bibinfo {author} {\bibfnamefont {Y.}~\bibnamefont
  {Yu}}\ and\ \bibinfo {author} {\bibfnamefont {K.}~\bibnamefont {Yang}},\
  }\href@noop {} {\bibfield  {journal} {\bibinfo  {journal} {Phys. Rev. Lett.}\
  }\textbf {\bibinfo {volume} {105}},\ \bibinfo {pages} {150605} (\bibinfo
  {year} {2010})}\BibitemShut {NoStop}%
\bibitem [{\citenamefont {Bauer}\ \emph {et~al.}(2013)\citenamefont {Bauer},
  \citenamefont {Huijse}, \citenamefont {Berg}, \citenamefont {Troyer},\ and\
  \citenamefont {Schoutens}}]{Bauer2013}%
  \BibitemOpen
  \bibfield  {author} {\bibinfo {author} {\bibfnamefont {B.}~\bibnamefont
  {Bauer}}, \bibinfo {author} {\bibfnamefont {L.}~\bibnamefont {Huijse}},
  \bibinfo {author} {\bibfnamefont {E.}~\bibnamefont {Berg}}, \bibinfo {author}
  {\bibfnamefont {M.}~\bibnamefont {Troyer}}, \ and\ \bibinfo {author}
  {\bibfnamefont {K.}~\bibnamefont {Schoutens}},\ }\href@noop {} {\bibfield
  {journal} {\bibinfo  {journal} {Phys. Rev. B}\ }\textbf {\bibinfo {volume}
  {87}},\ \bibinfo {pages} {165145} (\bibinfo {year} {2013})}\BibitemShut
  {NoStop}%
\bibitem [{\citenamefont {Huijse}\ \emph {et~al.}(2015)\citenamefont {Huijse},
  \citenamefont {Bauer},\ and\ \citenamefont {Berg}}]{Huijse2015}%
  \BibitemOpen
  \bibfield  {author} {\bibinfo {author} {\bibfnamefont {L.}~\bibnamefont
  {Huijse}}, \bibinfo {author} {\bibfnamefont {B.}~\bibnamefont {Bauer}}, \
  and\ \bibinfo {author} {\bibfnamefont {E.}~\bibnamefont {Berg}},\ }\href@noop
  {} {\bibfield  {journal} {\bibinfo  {journal} {Phys. Rev. Lett.}\ }\textbf
  {\bibinfo {volume} {114}},\ \bibinfo {pages} {090404} (\bibinfo {year}
  {2015})}\BibitemShut {NoStop}%
\bibitem [{\citenamefont {Jian}\ \emph {et~al.}(2015)\citenamefont {Jian},
  \citenamefont {Jiang},\ and\ \citenamefont {Yao}}]{Jian2015}%
  \BibitemOpen
  \bibfield  {author} {\bibinfo {author} {\bibfnamefont {S.-K.}\ \bibnamefont
  {Jian}}, \bibinfo {author} {\bibfnamefont {Y.-F.}\ \bibnamefont {Jiang}}, \
  and\ \bibinfo {author} {\bibfnamefont {H.}~\bibnamefont {Yao}},\ }\href@noop
  {} {\bibfield  {journal} {\bibinfo  {journal} {Phys. Rev. Lett.}\ }\textbf
  {\bibinfo {volume} {114}},\ \bibinfo {pages} {237001} (\bibinfo {year}
  {2015})}\BibitemShut {NoStop}%
\bibitem [{\citenamefont {Friedan}\ \emph {et~al.}(1984)\citenamefont
  {Friedan}, \citenamefont {Qiu},\ and\ \citenamefont {Shenker}}]{Friedan1984}%
  \BibitemOpen
  \bibfield  {author} {\bibinfo {author} {\bibfnamefont {D.}~\bibnamefont
  {Friedan}}, \bibinfo {author} {\bibfnamefont {Z.}~\bibnamefont {Qiu}}, \ and\
  \bibinfo {author} {\bibfnamefont {S.}~\bibnamefont {Shenker}},\ }\href@noop
  {} {\bibfield  {journal} {\bibinfo  {journal} {Phys. Rev. Lett.}\ }\textbf
  {\bibinfo {volume} {52}},\ \bibinfo {pages} {1575} (\bibinfo {year}
  {1984})}\BibitemShut {NoStop}%
\bibitem [{\citenamefont {Friedan}\ \emph {et~al.}(1985)\citenamefont
  {Friedan}, \citenamefont {Qiu},\ and\ \citenamefont {Shenker}}]{Friedan1985}%
  \BibitemOpen
  \bibfield  {author} {\bibinfo {author} {\bibfnamefont {D.}~\bibnamefont
  {Friedan}}, \bibinfo {author} {\bibfnamefont {Z.}~\bibnamefont {Qiu}}, \ and\
  \bibinfo {author} {\bibfnamefont {S.}~\bibnamefont {Shenker}},\ }\href@noop
  {} {\bibfield  {journal} {\bibinfo  {journal} {Phys. Lett. B}\ }\textbf
  {\bibinfo {volume} {151}},\ \bibinfo {pages} {37} (\bibinfo {year}
  {1985})}\BibitemShut {NoStop}%
\bibitem [{\citenamefont {Qiu}(1986)}]{Qiu1986}%
  \BibitemOpen
  \bibfield  {author} {\bibinfo {author} {\bibfnamefont {Z.}~\bibnamefont
  {Qiu}},\ }\href@noop {} {\bibfield  {journal} {\bibinfo  {journal} {Nucl.
  Phys. B}\ }\textbf {\bibinfo {volume} {270}},\ \bibinfo {pages} {205}
  (\bibinfo {year} {1986})}\BibitemShut {NoStop}%
\bibitem [{\citenamefont {Blume}(1966)}]{Blume1966}%
  \BibitemOpen
  \bibfield  {author} {\bibinfo {author} {\bibfnamefont {M.}~\bibnamefont
  {Blume}},\ }\href@noop {} {\bibfield  {journal} {\bibinfo  {journal} {Phys.
  Rev.}\ }\textbf {\bibinfo {volume} {141}},\ \bibinfo {pages} {517} (\bibinfo
  {year} {1966})}\BibitemShut {NoStop}%
\bibitem [{\citenamefont {Capel}(1966)}]{Capel1966}%
  \BibitemOpen
  \bibfield  {author} {\bibinfo {author} {\bibfnamefont {H.~W.}\ \bibnamefont
  {Capel}},\ }\href@noop {} {\bibfield  {journal} {\bibinfo  {journal} {Physica
  (Amsterdam)}\ }\textbf {\bibinfo {volume} {32}},\ \bibinfo {pages} {966}
  (\bibinfo {year} {1966})}\BibitemShut {NoStop}%
\bibitem [{\citenamefont {Alcaraz}\ \emph {et~al.}(1985)\citenamefont
  {Alcaraz}, \citenamefont {Drugowich~de Fel\'icio}, \citenamefont
  {K\"oberle},\ and\ \citenamefont {Stilck}}]{Alcaraz1985}%
  \BibitemOpen
  \bibfield  {author} {\bibinfo {author} {\bibfnamefont {F.~C.}\ \bibnamefont
  {Alcaraz}}, \bibinfo {author} {\bibfnamefont {J.~R.}\ \bibnamefont
  {Drugowich~de Fel\'icio}}, \bibinfo {author} {\bibfnamefont {R.}~\bibnamefont
  {K\"oberle}}, \ and\ \bibinfo {author} {\bibfnamefont {J.~F.}\ \bibnamefont
  {Stilck}},\ }\href@noop {} {\bibfield  {journal} {\bibinfo  {journal} {Phys.
  Rev. B}\ }\textbf {\bibinfo {volume} {32}},\ \bibinfo {pages} {7469}
  (\bibinfo {year} {1985})}\BibitemShut {NoStop}%
\bibitem [{\citenamefont {Zamolodchikov}(1991)}]{Zamolodchikov1991}%
  \BibitemOpen
  \bibfield  {author} {\bibinfo {author} {\bibfnamefont {A.~B.}\ \bibnamefont
  {Zamolodchikov}},\ }\href@noop {} {\bibfield  {journal} {\bibinfo  {journal}
  {Nucl. Phys. B}\ }\textbf {\bibinfo {volume} {358}},\ \bibinfo {pages} {524}
  (\bibinfo {year} {1991})}\BibitemShut {NoStop}%
\bibitem [{Note1()}]{Note1}%
  \BibitemOpen
  \bibinfo {note} {Right at the TCI point, SUSY is also spontaneously broken
  for a finite system with, say, periodic boundary conditions~\cite
  {Friedan1985}. The breaking of SUSY appears in $1/L$ corrections to the
  spectrum, while the operator content still reflects the superconformal
  structure of the CFT.}\BibitemShut {Stop}%
\bibitem [{\citenamefont {Grover}\ \emph {et~al.}(2014)\citenamefont {Grover},
  \citenamefont {Sheng},\ and\ \citenamefont {Vishwanath}}]{Grover2014}%
  \BibitemOpen
  \bibfield  {author} {\bibinfo {author} {\bibfnamefont {T.}~\bibnamefont
  {Grover}}, \bibinfo {author} {\bibfnamefont {D.~N.}\ \bibnamefont {Sheng}}, \
  and\ \bibinfo {author} {\bibfnamefont {A.}~\bibnamefont {Vishwanath}},\
  }\href@noop {} {\bibfield  {journal} {\bibinfo  {journal} {Science}\ }\textbf
  {\bibinfo {volume} {344}},\ \bibinfo {pages} {6181} (\bibinfo {year}
  {2014})}\BibitemShut {NoStop}%
\bibitem [{\citenamefont {Kastor}\ \emph {et~al.}(1989)\citenamefont {Kastor},
  \citenamefont {Martinec},\ and\ \citenamefont {Shenker}}]{Kastor1989}%
  \BibitemOpen
  \bibfield  {author} {\bibinfo {author} {\bibfnamefont {D.~A.}\ \bibnamefont
  {Kastor}}, \bibinfo {author} {\bibfnamefont {E.~J.}\ \bibnamefont
  {Martinec}}, \ and\ \bibinfo {author} {\bibfnamefont {S.~H.}\ \bibnamefont
  {Shenker}},\ }\href@noop {} {\bibfield  {journal} {\bibinfo  {journal} {Nucl.
  Phys. B}\ }\textbf {\bibinfo {volume} {316}},\ \bibinfo {pages} {590}
  (\bibinfo {year} {1989})}\BibitemShut {NoStop}%
\bibitem [{Note2()}]{Note2}%
  \BibitemOpen
  \bibinfo {note} {The sign of the hopping amplitude is unimportant as it can
  be changed via the transformation $\gamma _{j}\to (-1)^j\gamma
  _{j}$.}\BibitemShut {Stop}%
\bibitem [{\citenamefont {Milsted}\ \emph {et~al.}()\citenamefont {Milsted},
  \citenamefont {Seabra}, \citenamefont {Fulga}, \citenamefont {Beenakker},\
  and\ \citenamefont {Cobanera}}]{Milsted2015}%
  \BibitemOpen
  \bibfield  {author} {\bibinfo {author} {\bibfnamefont {A.}~\bibnamefont
  {Milsted}}, \bibinfo {author} {\bibfnamefont {L.}~\bibnamefont {Seabra}},
  \bibinfo {author} {\bibfnamefont {I.~C.}\ \bibnamefont {Fulga}}, \bibinfo
  {author} {\bibfnamefont {C.~W.~J.}\ \bibnamefont {Beenakker}}, \ and\
  \bibinfo {author} {\bibfnamefont {E.}~\bibnamefont {Cobanera}},\ }\href@noop
  {} {}\Eprint {http://arxiv.org/abs/1504.07258} {arXiv:1504.07258}
  \BibitemShut {NoStop}%
\bibitem [{\citenamefont {Rahmani}\ \emph {et~al.}()\citenamefont {Rahmani},
  \citenamefont {Zhu}, \citenamefont {Franz},\ and\ \citenamefont
  {Affleck}}]{Rahmani2015}%
  \BibitemOpen
  \bibfield  {author} {\bibinfo {author} {\bibfnamefont {A.}~\bibnamefont
  {Rahmani}}, \bibinfo {author} {\bibfnamefont {X.}~\bibnamefont {Zhu}},
  \bibinfo {author} {\bibfnamefont {M.}~\bibnamefont {Franz}}, \ and\ \bibinfo
  {author} {\bibfnamefont {I.}~\bibnamefont {Affleck}},\ }\href@noop {}
  {}\Eprint {http://arxiv.org/abs/1505.03966} {arXiv:1505.03966} \BibitemShut
  {NoStop}%
\bibitem [{\citenamefont {Teo}\ and\ \citenamefont {Kane}(2010)}]{Teo2010}%
  \BibitemOpen
  \bibfield  {author} {\bibinfo {author} {\bibfnamefont {J.~C.~Y.}\
  \bibnamefont {Teo}}\ and\ \bibinfo {author} {\bibfnamefont {C.~L.}\
  \bibnamefont {Kane}},\ }\href@noop {} {\bibfield  {journal} {\bibinfo
  {journal} {Phys. Rev. B}\ }\textbf {\bibinfo {volume} {82}},\ \bibinfo
  {pages} {115120} (\bibinfo {year} {2010})}\BibitemShut {NoStop}%
\bibitem [{Note3()}]{Note3}%
  \BibitemOpen
  \bibinfo {note} {This is convenient because translation symmetry is
  spontaneously broken at strong coupling.}\BibitemShut {Stop}%
\bibitem [{Note4()}]{Note4}%
  \BibitemOpen
  \bibinfo {note} {See Supplemental Material [url], which includes Ref. \cite
  {Mahan}}\BibitemShut {NoStop}%
  \bibitem{Mahan} G.D. Mahan, \textit{Many-Particle Physics}, Ch.\ 9.3  (Plenum Press, New York) 1981.
\bibitem [{\citenamefont {Selke}(1988)}]{Selke1988}%
  \BibitemOpen
  \bibfield  {author} {\bibinfo {author} {\bibfnamefont {W.}~\bibnamefont
  {Selke}},\ }\href@noop {} {\bibfield  {journal} {\bibinfo  {journal} {Phys.
  Rep.}\ }\textbf {\bibinfo {volume} {170}},\ \bibinfo {pages} {213} (\bibinfo
  {year} {1988})}\BibitemShut {NoStop}%
\bibitem [{\citenamefont {Turban}(1982)}]{Turban1982}%
  \BibitemOpen
  \bibfield  {author} {\bibinfo {author} {\bibfnamefont {L.}~\bibnamefont
  {Turban}},\ }\href@noop {} {\bibfield  {journal} {\bibinfo  {journal} {J.
  Phys. C}\ }\textbf {\bibinfo {volume} {15}},\ \bibinfo {pages} {L65}
  (\bibinfo {year} {1982})}\BibitemShut {NoStop}%
\bibitem [{\citenamefont {Penson}\ \emph {et~al.}(1982)\citenamefont {Penson},
  \citenamefont {Jullien},\ and\ \citenamefont {Pfeuty}}]{Penson1982}%
  \BibitemOpen
  \bibfield  {author} {\bibinfo {author} {\bibfnamefont {K.~A.}\ \bibnamefont
  {Penson}}, \bibinfo {author} {\bibfnamefont {R.}~\bibnamefont {Jullien}}, \
  and\ \bibinfo {author} {\bibfnamefont {P.}~\bibnamefont {Pfeuty}},\
  }\href@noop {} {\bibfield  {journal} {\bibinfo  {journal} {Phys. Rev.}\
  }\textbf {\bibinfo {volume} {26}},\ \bibinfo {pages} {6334} (\bibinfo {year}
  {1982})}\BibitemShut {NoStop}%
\bibitem [{\citenamefont {Alcaraz}(1986)}]{Alcaraz1986}%
  \BibitemOpen
  \bibfield  {author} {\bibinfo {author} {\bibfnamefont {F.~C.}\ \bibnamefont
  {Alcaraz}},\ }\href@noop {} {\bibfield  {journal} {\bibinfo  {journal} {Phys.
  Rev. B}\ }\textbf {\bibinfo {volume} {34}},\ \bibinfo {pages} {4885}
  (\bibinfo {year} {1986})}\BibitemShut {NoStop}%
\bibitem [{\citenamefont {Bl\"ote}\ \emph {et~al.}(1986)\citenamefont
  {Bl\"ote}, \citenamefont {Compagner}, \citenamefont {Cornelissen},
  \citenamefont {Hoogland}, \citenamefont {Mallezie},\ and\ \citenamefont
  {Vanderzande}}]{Blote1986}%
  \BibitemOpen
  \bibfield  {author} {\bibinfo {author} {\bibfnamefont {H.}~\bibnamefont
  {Bl\"ote}}, \bibinfo {author} {\bibfnamefont {A.}~\bibnamefont {Compagner}},
  \bibinfo {author} {\bibfnamefont {P.}~\bibnamefont {Cornelissen}}, \bibinfo
  {author} {\bibfnamefont {A.}~\bibnamefont {Hoogland}}, \bibinfo {author}
  {\bibfnamefont {F.}~\bibnamefont {Mallezie}}, \ and\ \bibinfo {author}
  {\bibfnamefont {C.}~\bibnamefont {Vanderzande}},\ }\href@noop {} {\bibfield
  {journal} {\bibinfo  {journal} {Physica A}\ }\textbf {\bibinfo {volume}
  {139}},\ \bibinfo {pages} {395} (\bibinfo {year} {1986})}\BibitemShut
  {NoStop}%
\bibitem [{\citenamefont {Cardy}(1996)}]{Cardy}%
  \BibitemOpen
  \bibfield  {author} {\bibinfo {author} {\bibfnamefont {J.}~\bibnamefont
  {Cardy}},\ }\href@noop {} {\emph {\bibinfo {title} {Scaling and
  Renormalization in Statistical Physics}}}\ (\bibinfo  {publisher} {Cambridge
  University Press},\ \bibinfo {year} {1996})\BibitemShut {NoStop}%
\bibitem [{Note5()}]{Note5}%
  \BibitemOpen
  \bibinfo {note} {We can accurately extract $E^{\protect \rm even}_{{\protect
  \rm A}, 0}-\epsilon _0 L$ from the numerics by a linear fit of the energy
  density $E^{\protect \rm even}_{{\protect \rm A}, 0}/L$ to $1/
  L^2$.}\BibitemShut {Stop}%
\bibitem [{\citenamefont {Cappelli}(1987)}]{Cappelli1987}%
  \BibitemOpen
  \bibfield  {author} {\bibinfo {author} {\bibfnamefont {A.}~\bibnamefont
  {Cappelli}},\ }\href@noop {} {\bibfield  {journal} {\bibinfo  {journal}
  {Phys. Lett. B}\ }\textbf {\bibinfo {volume} {185}},\ \bibinfo {pages} {82}
  (\bibinfo {year} {1987})}\BibitemShut {NoStop}%
\bibitem [{\citenamefont {L\"assig}\ \emph {et~al.}(1991)\citenamefont
  {L\"assig}, \citenamefont {Mussardo},\ and\ \citenamefont
  {Cardy}}]{Lassig1991}%
  \BibitemOpen
  \bibfield  {author} {\bibinfo {author} {\bibfnamefont {M.}~\bibnamefont
  {L\"assig}}, \bibinfo {author} {\bibfnamefont {G.}~\bibnamefont {Mussardo}},
  \ and\ \bibinfo {author} {\bibfnamefont {J.~L.}\ \bibnamefont {Cardy}},\
  }\href@noop {} {\bibfield  {journal} {\bibinfo  {journal} {Nucl. Phys. B}\
  }\textbf {\bibinfo {volume} {348}},\ \bibinfo {pages} {594} (\bibinfo {year}
  {1991})}\BibitemShut {NoStop}%
\bibitem [{\citenamefont {Cardy}(1986)}]{Cardy1986}%
  \BibitemOpen
  \bibfield  {author} {\bibinfo {author} {\bibfnamefont {J.~L.}\ \bibnamefont
  {Cardy}},\ }\href@noop {} {\bibfield  {journal} {\bibinfo  {journal} {Nucl.
  Phys. B}\ }\textbf {\bibinfo {volume} {270}},\ \bibinfo {pages} {186}
  (\bibinfo {year} {1986})}\BibitemShut {NoStop}%
\bibitem [{\citenamefont {Cheng}\ \emph {et~al.}(2009)\citenamefont {Cheng},
  \citenamefont {Lutchyn}, \citenamefont {Galitski},\ and\ \citenamefont
  {Das~Sarma}}]{Cheng2009}%
  \BibitemOpen
  \bibfield  {author} {\bibinfo {author} {\bibfnamefont {M.}~\bibnamefont
  {Cheng}}, \bibinfo {author} {\bibfnamefont {R.~M.}\ \bibnamefont {Lutchyn}},
  \bibinfo {author} {\bibfnamefont {V.}~\bibnamefont {Galitski}}, \ and\
  \bibinfo {author} {\bibfnamefont {S.}~\bibnamefont {Das~Sarma}},\ }\href@noop
  {} {\bibfield  {journal} {\bibinfo  {journal} {Phys. Rev. Lett.}\ }\textbf
  {\bibinfo {volume} {103}},\ \bibinfo {pages} {107001} (\bibinfo {year}
  {2009})}\BibitemShut {NoStop}%
\end{thebibliography}%

\end{document}

\title{Supplemental Material for ``Emergent Supersymmetry from Strongly Interacting Majorana Fermions''}
\maketitle

\section{Connection to spin chains}
A standard Jordan-Wigner transformation $\sigma_j^z=2n_j-1$ and $\sigma_j^+=e^{i\pi\sum_{k<j} n_k}c^\dagger_j$ brings the Hamiltonian~(2) of the main text to the form of a spin chain
 \begin{equation}\label{eq:spin_chain}
H=t_1\sum_j\sigma_j^z-t_2\sum_j\sigma_j^x \sigma_{j+1}^x-g_1\sum_j\sigma_j^z \sigma_{j+1}^z-g_2\sum_j\sigma_j^x \sigma_{j+2}^x,
\end{equation}
which for $g_1=0$  corresponds to the anisotropic next-nearest-neighbor Ising (ANNNI) model~\cite{Selke1988}.
For $t_1=t_2=0$,
 the Hamiltonian~(\ref{eq:spin_chain}) is dual to a transverse field Ising model with 4-spin interactions
 \begin{equation}\label{eq:spin_chain_dual}
H_g=-g_1\sum_j\tau_j^z-g_2\sum_j\tau_{j-1}^x \tau_{j}^x\tau_{j+1}^x\tau_{j+2}^x,
\end{equation}
in terms of domain-wall variable $\tau_j^z=
\sigma_j^z\sigma_{j+1}^z$.

  The Hamiltonian~(\ref{eq:spin_chain_dual}) is self-dual at $g_1=g_2$~\cite{Turban1982,Penson1982}. {As argued in Refs.~\cite{Alcaraz1986,Blote1986}, the ground state of $H_g$ is 8-fold degenerate for $g_1=0$, while the $g_1$ term causes quantum fluctuations between these degenerate states. This bears a striking similarity to the 8-state Potts model, which strongly suggests a first-order phase transition at $g_1=g_2$~\cite{Alcaraz1986,Blote1986}. }

\section{Similarities with the Blume-Capel model}  
As mentioned in the main text, the canonical example of the TCI CFT appears in the Blume-Capel model, which is a quantum spin-1 chain with the Hamiltonian
\begin{equation}
H_{\rm BC}=-\sum_j\left[S_z(j) S_z(j+1)-\gamma S_x(j)-\delta S^2_z(j)\right].
\end{equation}
Similar to the one-dimensional spin-$1\over 2$ transverse-field Ising model, which corresponds to the two-dimensional classical Ising model at finite temperature, this spin chain corresponds to a two-dimensional classical Ising model with vacancies ($S_z=0$), with $\delta$ serving as a chemical potential for these vacancies. For $\delta=-\infty$, the Blume-Capel model maps to the transverse-filed Ising model and exhibits a 2nd-order phase transition between a ferromagnet and a paramagnet with the critical point given by the Ising CFT. For $\gamma=0$, the model is classical and  has a 1st-order transition between a ferromagnet and the $S_z(j)=0$ state. The line of the 2nd-order transitions terminates at the TCI point, where a line of 1st-order transitions begins.

The above Hamiltonian has two dimensionless parameters. We can write a generalization of our model, which also has two dimensionless parameters, by staggering the hopping terms [Eq. (2) of the main text with $g_1=g_2=g>0$]. The phase diagrams are essentially the same as seen in Fig. 1. In our model broken translation $t_1-t_2\neq 0$ immediately gaps the system.
\begin{figure}[center]
	\includegraphics[width=7.5cm]{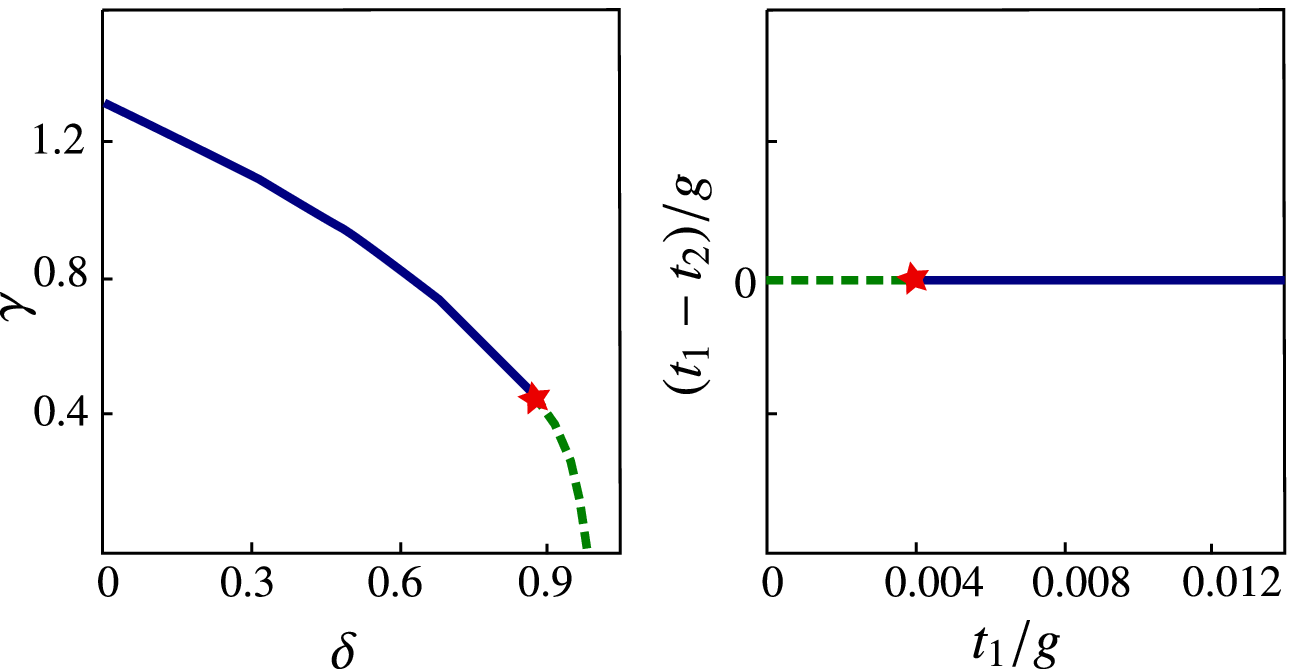}
	\caption{Left: the phase diagram of the Blume-Capel model (3) from Ref.~\cite{Alcaraz1985}. Right: the phase diagram of our model with staggered hopping [Eq.(2) of the main text] with $g_1=g_2=g>0$. The solid blue (dashed green) line represents a 2nd (1st) order transition with the TCI point denoted by a red star.}
\end{figure}

\section{finite size spectrum from CFT}

The finite size spectrum of the Ising and TCI model with both APBC and PBC can be derived using modular invariance~\cite{Cardy1986}. Consider an imaginary time Feynman path integral for the partition function 
at inverse temperature $\beta$ for free relativistic Majorana fermions.  As is well known, 
the boundary conditions must be antiperiodic in the imaginary time direction. Thus, when we also impose APBC in the space 
direction the partition function 
\be Z=\sum_ne^{-2\pi vy_n\beta /L}, \text{ with }E_n={2\pi vy_n\over L}\ee
is modular invariant.
Each state corresponds to a left and right-moving factor and the energy is a sum of left and right moving parts:
\be y_n=x_n+\bar x_n, \text{ with }x_n=-{c\over 24}+x_i+n,\ee
for nonnegative integers $n$, constants $x_i$, which characterize each conformal tower of states, and the central charge $c$ of the CFT. Each excited state in left and right moving conformal towers can occur independently. We define the character corresponding to a given chiral conformal tower as
\be \chi_i(\beta /L)\equiv \sum_{n=0}^\infty d_{in}e^{-(2\pi v\beta /L)(-c/24+x_i+n)},\ee
where $d_{in}$ are nonnegative integers, which account for possible degeneracies of excited states in each conformal tower. 
The partition function then has the form
\be Z(\beta /L)=\sum_{i,j}N_{ij}\chi_i(\beta /L)\chi_j(\beta /L)\label{Z}.\ee
Here the $N_{ij}$ are nonnegative integer encoding how many times each possible product of left and right conformal towers occurs. 
Modular invariance imposes a restriction on the multiplicities $N_{ij}$~\cite{Cardy1986}
\be N_{ij}=\sum_{i'j'}S_{ii'}S_{jj'}N_{i'j'},\label{modc}\ee
where $S$ is the modular S-matrix.

For the Ising CFT, there are three conformal towers
\be(\sigma ,\epsilon ,I )\leftrightarrow(1,2,3)\ee
 with the corresponding values of $x_i$ given by $(1/16,1/2,0)$.
The Ising modular S-matrix is
\be S=\left(\begin{array}{ccc} 0&-1/\sqrt{2}&1/\sqrt{2}\\ -1/\sqrt{2}&1/2&1/2\\ 1/\sqrt{2}&1/2&1/2 \end{array}\right).\ee

{There are two realizations of the Ising CFT with different local operator content, a spin model and a fermionic model. In the spin model, all three primary fields are present as local operators. A solution of Eq.~(\ref{modc}) for the multiplicities is given by $N_{ij}=\delta_{ij}$, which leads to the following operator content:
\be
(I,I), \quad (\sigma,\sigma), \quad (\epsilon, \epsilon).
\ee
As spins are bosonic, the partition function with periodic boundary conditions corresponds to the local operator content.
For the fermionic model, the field $\sigma$, which corresponds to the Ising order parameter, does not appear in the operator content since this is nonlocal in terms of fermions. Eq.~(\ref{modc}) then has a unique solution as eliminating the first row and column corresponding to $\sigma$ leads to the eigenvalue equation 
\be\left(\begin{array}{cc} 1/2&1/2\\ 1/2&1/2 \end{array}\right)\otimes\left(\begin{array}{cc} 1/2&1/2\\ 1/2&1/2 \end{array}\right)|N\rangle =|N\rangle.\ee
The matrix on the left-hand side has only one eigenvector $\langle N|=(1,1,1,1)$ with eigenvalue 1, which gives the operator content
\be
(I,I), \quad (I,\epsilon),\quad (\epsilon, I),\quad (\epsilon, \epsilon).
\ee

For fermions, the partition function with antiperiodic boundary conditions corresponds to the local operator content and can be written as}
\be Z_{A,A}=\chi_2^2+\chi_3^2+2\chi_2\chi_3.\label{ZAAI}\ee
The spectrum with PBC for the fermionic model is given by $Z_{A,P}$,
where the first subscript refers to time and the second to space. $Z_{A,P}$ can be determined by a modular transformation from $Z_{P,A}$, where
\be Z_{P,A}={\rm tr}(-1)^Fe^{-\beta H},\label{ZPA}\ee
{with $F$ denoting the fermion parity. Notice that the operators $(I,\epsilon)$ and $(\epsilon, I)$ have conformal spin $\mp {1\over 2}$ and are therefore fermionic.} To obtain $Z_{P,A}$, we simply insert a minus sign for all states with odd fermion parity. Thus
\be Z_{P,A}=\chi_2^2+\chi_3^2-2\chi_2\chi_3.\ee
Since $Z_{A,P}$ is related to $Z_{P,A}$ by 
\be Z_{A,P}=\sum_{i,j,i',j'} \chi_i\chi_jS_{ii'}S_{jj'}N_{i',j'}^{(P,A)},\ee
we conclude that
\be Z_{A,P}=2\chi_1^2.\label{ZAPI}\ee

We can now extend this to the TCI model~\cite{Friedan1984}. For the TCI CFT, there are 6 chiral conformal towers 
\be(\epsilon ,\epsilon ',\epsilon '',\sigma ,\sigma ',I)\leftrightarrow (1,2,3,4,5,6),\ee
 with $(1/10,3/5,3/2,3/80,7/16,0)$ being the corresponding $x_i$.  {The modular S matrix of the TCI model is given by
\[ S={1\over \sqrt{5}}\left(\begin{array}{cccccc} -s_1&-s_1&s_2&\sqrt{2}s_1&-\sqrt{2}s_2&s_2\\
-s_1&-s_1&s_2&-\sqrt{2}s_1&\sqrt{2}s_2&s_2\\
s_2&s_2&s_1&-\sqrt{2}s_2&-\sqrt{2}s_1&s_1\\
\sqrt{2}s_1&-\sqrt{2}s_1&-\sqrt{2}s_2&0&0&\sqrt{2}s_2\\
-\sqrt{2}s_2&\sqrt{2}s_2&-\sqrt{2}s_1&0&0&\sqrt{2}s_1\\
s_2&s_2&s_1&\sqrt{2}s_2&\sqrt{2}s_1&s_1
\end{array}\right),\] 
 where $s_j\equiv \sin (j\pi /5)$. Once again, we have a spin model and a fermionic model. Similar to the Ising case, the multiplicities  of the spin model are given by $N_{ij}=\delta_{ij}$ due to modular invariance. 
 We can then write the operator content of the spin model as
\be
\underbrace{(I,I),\;(\epsilon,\epsilon),\; (\epsilon',\epsilon'),\;(\epsilon'',\epsilon'')}_{{\rm local~in~spin~and~fermion~models},~Z_2~{\rm even}},\:\underbrace{(\sigma,\sigma),\;(\sigma',\sigma').}_{{\rm local~only~in~spin~model},~Z_2~{\rm odd}}
\ee
 Similar to the Ising case, the operator content of the fermionic TCI model can be obtained form modular invariance and requiring that the nonlocal (in fermions) operators $\sigma$ and $\sigma'$ do not appear. Again, this leads to a unique solution to Eq.~(\ref{modc}) with the following local operator content for the fermionic model:
 \be
\underbrace{(I,I),\;(\epsilon,\epsilon),\; (\epsilon',\epsilon'),\;(\epsilon'',\epsilon'')}_{{\rm local~in~spin~and~fermion~models},~Z_2~{\rm even}},\:\underbrace{(\epsilon,\epsilon'),\;(\epsilon',\epsilon),\:(I,\epsilon''),\;(\epsilon'',I).}_{{\rm local~only~in~fermion~model},~Z_2~{\rm even}}
\ee
%We note in passing that the Hilbert space of the TCI model consists of two sectors known as the Neveu-Schwartz and the Ramond sectors. Of all the operators above (in both spin and fermion models), only the $Z_2$-odd operators $(\sigma,\sigma)$ and $(\sigma',\sigma')$, which only appear in the spin model, have matrix elements between these sectors and the other operators are diagonal with respect to the two sectors.  

As in the Ising case, the local operator content of the fermionic model corresponds to the modular invariant spectrum with antiperiodic boundary conditions~\cite{Cappelli1987}:
\be Z_{A,A}=(\chi_1+\chi_2)^2+(\chi_3+\chi_6)^2.\ee
Noting that operators $(\epsilon',\epsilon)$ and $(\epsilon'',I)$ respectively have half-odd integer conformal spins $3/5-1/10=1/2$ and $3/2-0=3/2$ (and are thus fermionic), we can write $Z_{P,A}=(\chi_1-\chi_2)^2+(\chi_3-\chi_6)^2$ upon the insertion of $(-1)^F$. From Eq. (14), we can then write the partition function with periodic boundary conditions as
\begin{eqnarray}
Z_{A,P}&=&2(\chi_4^2+\chi_5^2).
\end{eqnarray}
}
From the above equations, we can read off the complete spectrum with APBC and PBC for both Ising and TCI models. Note that the states with odd/even fermion parity correspond to the fermonic/bosonic operators $\chi_i$.

For Ising model, the ground state with APBC has $x_i=\bar x_i=0$ and is in the even sector. The degenerate ground states with PBC have $x_i+\bar x_i=1/16+1/16=1/8$ and thus have energy higher by $(2\pi v/L)(1/8)$ than the unique ground state with APBC. The first excited states with PBC are a degenerate pair with energy $(2\pi v/L)(1/4)$. For TCI model, the unique ground state with APBC also has $x_i=\bar x_i=0$. With PBC, the degenerate ground states have energy higher by $(2\pi v/L)(3/40)$ than that with APBC, and the first excited states with PBC have energy $(2\pi v/L)(7/8)$.

%%%%%%%%%

\section{Tunneling Density of states}
Modular invariance also implies a relationship between the scaling dimensions of 
operators determining the exponents on their power-law decaying correlation functions and the parameters $x_n$, $\bar x_n$ in the 
finite size energies. To each state in the spectrum with APBC there is a corresponding operator $O_{x_n,\bar x_n}$ with $\langle O_{x_n,\bar x_n}(\tau ,x)O_{x_n,\bar x_n}(0,0)\rangle\propto {1\over (\tau +ix)^{2x_n}(\tau -ix)^{2\bar x_n}}$. We use this general relation to determine the scaling of tunneling current with voltage in scanning tunneling experiments.

Now let's consider the single fermion Green's function. For the free Majorana model (Ising model), we have two chiral operators $(\epsilon ,I)$ and $(I ,\epsilon)$, corresponding to Majorana operators $\gamma_R$ and $\gamma_L$, where $\gamma_{R,L}=(\alpha\pm\beta)/2$. The only correlations are then
\be \langle\gamma_{R,L}(t,x),\gamma_{R,L}(0,0)\rangle={i\over 4\pi (vt\pm x+i\delta )},\ee
where $\delta$ is a positive constant of order the lattice spacing (one).  
Note that at equal times 
\bea \langle \alpha_j\alpha_0\rangle &=&\langle \beta_j\beta_0\rangle =0\nonumber \\
\langle \alpha_j\beta_0\rangle &=&\langle \beta_j\alpha_0\rangle ={i\over 2\pi j}.\label{ET}
\eea
On the other hand, for $j=0$ but unequal times,
\be \langle \alpha_0(t)\alpha_0(0)\rangle =\langle \beta_0(t)\beta_0(0)\rangle ={i\over 2\pi (vt+i\delta )}\ee 

In STM we are interested in the tunneling current between the sample and the normal tip, given by \cite{mahan1}
\be \langle I\rangle
=-2e^2\Gamma^2\hbox{Im}G_R(-eV)\ee
where the retarded Green's function
\be G_R(\omega)=-i\int_0^\infty dt e^{i\omega t}\langle[\alpha_0(t)\psi_0(t),\alpha_0(0)\psi^\dagger_0(0)]\rangle\ee
is given in terms of Majorana fields $\alpha_0$ and ordinary fermions $\psi_0$ describing the tip.
It is convenient to obtain this retarded Green's function by analytic continuation from the Matsubara Green's function
$ G_R(\omega )=G_M(i\omega_n\to \delta +i\omega )$
which conveniently  factorizes 
\be G_M(\tau )=-\langle T\alpha_0(\tau )\alpha_0(0)\rangle \langle T\psi_0(\tau )\psi_0^\dagger (0)\rangle\label{GM},\ee
where $T$ indicates time ordering. 
The Green's function for the normal region at zero temperature is
$ G_N(\tau )\approx -{D_N/ \tau }$ with $D_N$ the density of states.
For the noninteracting limit of the Majorana model
$ -\langle T\alpha_0(\tau )\alpha_0(0)\rangle\approx -{1/2\pi  v\tau}$
thus obtaining
\be G_M(\tau )\approx {D_N\over 2\pi v\tau^2}.\ee
Performing the Fourier transform followed by the analytical continuation we find $\hbox{Im}G_R(-eV)=(D_N/ v)eV$.
This gives a tunneling current
\be I=-2\pi \Gamma^2e^2{D_N\over v}eV\label{I}.\ee

Now consider the TCI model. We expect that 
  \be \chi_R\to c_1\chi_{R1}+c_2\chi_{R2}\ee
where $\chi_{Ri}$ are the two fermonic primary fields of dimension $(x,\bar x)=(3/5,1/10)$ and $(3/2,0)$ respectively.  The $c_i$ 
are nonuniversal constants. The corresponding correlation functions are
\bea \langle \chi_{R1}(t,x)\chi_{R1}(0,0)\rangle &=&{i\over2\pi  (vt-x+i\delta )[(vt+i\delta )^2-x^2]^{1/5}}\nonumber \\
\langle \chi_{R2}(t,x)\chi_{R2}(0,0)\rangle &=&{i\over2\pi  (vt-x+i\delta )^3}.\eea
The Green's functions for $\chi_{Lj}$ are obtained by taking $x\to -x$.  The $\chi_1$ term in $\chi$ should dominate at long times 
and low frequencies, so at equal times
\be \langle \alpha_j\beta_0\rangle \propto {\epsilon (j)\over |j|^{7/5}}.\ee
Using the Matsubara Green's function
\be \langle T\alpha_0(\tau )\alpha_0(0)\rangle \propto {1\over \tau |\tau |^{2/5}},\ee
and following the same set of steps as in the Ising case we find
\be \hbox{Im}G_R(\omega )\propto C\hbox{sign}(\omega )|\omega |^{7/5}\sin (7\pi /10),\ee
where $C$ is a positive real constant which can be expressed in terms of the $\Gamma$ functions. The tunneling current, therefore reads
\be I\propto \hbox{sign}(eV)|eV|^{7/5}.\ee

Another quantity that can be predicted is how the crossover scale, from tri-critical Ising  behavior, depends on $t-t_c$. The leading bosonic operator consistent with the 
spatial parity symmetry is $(\epsilon ',\epsilon ')$ of dimension 6/5, $O_{6/5}$. The effective 
Hamiltonian density near $t_c$ 
\be {\cal H}={\cal H}_{TC}+A(t-t_c)O_{6/5}\ee
where $A$ is a constant of dimension (length)$^{6/5}$. Under renormalization, this 
coupling constant scales as $L^{2-6/5}=L^{4/5}$. Thus we can define a crossover length scale
\be \xi \propto |t-t_c|^{-5/4}.\ee
For $t<t_c$, in the gapped phase, we should expect the gap to scale as
\be \Delta \propto (t_c-t)^{5/4}.\ee

%...